\documentclass[12pt]{article}

\usepackage{amsmath,amssymb,amsthm}

\usepackage{xchicago}
\makeatletter

\def\citeAPY{\def\@citeseppen{-1000}%
    \def\@cite##1##2{##1\if@tempswa , ##2\fi}%
    \def\citeauthoryear##1##2##3{##2 (##3)}\@internalcite}

\makeatother
\newcommand {\citeAY} [1] {\citeNP {#1}}%
\renewcommand {\showoriginalref}[1]{}
\renewcommand {\showCODEN}[1]{}
\renewcommand {\showISSN}[1]{}
\renewcommand {\showMR}[3]{}

\newcommand\eq[1] {(\ref{#1})}

\newcommand\fig[1] {\ref{fig:#1}}

\newcommand\labfig[1] {\label{fig:#1}}

\newcommand{\nonum}{\nonumber \\}
\newcommand{\beqa}{\begin{eqnarray}}
\newcommand{\eeqa}[1]{\label{#1}\end{eqnarray}}
\newcommand{\beq}{\begin{equation}}
\newcommand{\eeq}[1]{\label{#1}\end{equation}}

\newcommand{\Ga}{\alpha}
\newcommand{\Gb}{\beta}

\newcommand{\Ge}{\epsilon}
\newcommand{\Gve}{\varepsilon}
\newcommand{\Gf}{\phi}
\newcommand{\Gvf}{\varphi}
\newcommand{\Gg}{\gamma}

\newcommand{\Gl}{\lambda}
\newcommand{\Gn}{\eta}

\newcommand{\Gv}{\nu}

\newcommand{\Gt}{\theta}

\newcommand{\Gj}{\tau}

\newcommand{\CA}{{\cal A}}
\newcommand{\CB}{{\cal B}}

\newcommand{\CD}{{\cal D}}

\newcommand{\CP}{{\cal P}}

\newcommand{\CR}{{\cal R}}

\def\Bb{{\bf b}}

\def\Bt{{\bf t}}
\def\Bu{{\bf u}}
\def\Bv{{\bf v}}
\def\Bw{{\bf w}}
\def\Bx{{\bf x}}

\def\BA{{\bf A}}
\def\BB{{\bf B}}
\def\BC{{\bf C}}

\def\BF{{\bf F}}

\def\BI{{\bf I}}

\def\BM{{\bf M}}

\def\BR{{\bf R}}

\def \ba {\begin{array}}
\def \ea {\end{array}}

\def \refe #1.{(\ref{#1})}
\def \reff #1.{figure~\ref{#1}}
\def \refs #1.{section~\ref{#1}}
\def \refss #1.{subsection~\ref{#1}}
\def \refD #1.{Definition~\ref{#1}}
\def \refT #1.{Theorem~\ref{#1}}
\def \refL #1.{Lemma~\ref{#1}}
\def \refC #1.{Corollary~\ref{#1}}
\def \refP #1.{Proposition~\ref{#1}}
\def \refR #1.{Remark~\ref{#1}}
\def \refE #1.{Example~\ref{#1}}
\def \refN #1.{Notation~\ref{#1}}

\usepackage{graphics}
\usepackage{amssymb}
\begin{document}
\vspace{-1in}
\title{Complete characterization of the macroscopic deformations of periodic unimode metamaterials of rigid bars and pivots.}
\author{Graeme Walter Milton\\
\small{Department of Mathematics, University of Utah, Salt Lake City UT 84112, USA}}
\date{}
\maketitle
\begin{abstract}
A complete characterization is given of the possible macroscopic deformations of periodic nonlinear affine unimode 
metamaterials constructed from rigid bars and pivots. The materials are affine in 
the sense that their macroscopic deformations can only be affine deformations: on a local level the deformation may vary 
from cell to cell. Unimode means that macroscopically
the material can only deform along a one dimensional trajectory in the six dimensional space of invariants
describing the deformation (excluding translations and rotations). We show by explicit construction that 
any continuous trajectory is realizable to an arbitrarily high degree of approximation provided at all points along the trajectory the 
geometry does not collapse to a lower dimensional one. In particular, we present two and three dimensional dilational materials having an arbitrarily large flexibility window. These are perfect auxetic materials for which a dilation is the only easy mode of deformation. They are free to dilate to arbitrarily large strain with zero bulk modulus.
\end{abstract}
\vskip2mm
\noindent Keywords: A. microstructures, B. inhomogeneous material; B rods and cables; B foam material; unimode material  
\noindent 
\vskip2mm
\section{Introduction}
\setcounter{equation}{0} 

\citeAPY{Milton:1995:WET} (see also chapter 30 of \citeAPY{Milton:2002:TOC}) introduced the concept of unimode, bimode, trimode, quadramode and 
pentamode materials in the context of linear elasticity. In these extremal 
three-dimensional materials
the six eigenvalues of the elasticity matrix split into two groups: very large ones and very small ones: the number of very small ones (counting
degeneracy) determines the category of extremal material: unimode if there is one small eigenvalue, bimode if there are two, trimode if there are three,
quadramode if there are four, pentamode if there are five. For periodic pin-jointed trusses the category, and the infinitesimal modes of 
deformation, can be determined by an application of Bloch's theorem as noted by \citeAPY{Hutchinson:2006:SPP}. 

An example of a unimode material is an elastically isotropic material with a Poisson's ratio arbitrarily close to
$-1$ so that the only easy mode of deformation is an infinitesimal dilation. Isotropic materials with negative Poisson's ratio, the existence of which
was questioned for a long time, were first manufactured by \citeAPY{Lakes:1987:FSN}. Elastically isotropic materials with a Poisson's ratio arbitrarily close to $-1$ were 
found by \citeAPY{Almgren:1985:ITD} and \citeAPY{Milton:1992:CMP}. The constructions of Almgren used sliding collars, 
whereas the constructions of Milton did not. 
There has since been extensive work on materials with negative 
Poisson's ratio which are known as Auxetic materials: see, for example \citeAPY{Greaves:2011:PRM} and \citeAPY{Mitschke:2011:FAF} and references therein. 
An example of an isotropic pentamode material is a gel: it is easy to deform 
under any five independent infinitesimal shears, yet strongly resists hydrostatic compression. In general, pentamode materials can resist a desired combination of compression and shear,
and as such  they are useful for transformation acoustics and in particular cloaking for acoustics, as discovered by \citeAPY{Norris:2008:ACT}. Designs for these more general
 pentamode materials suggested by \citeAPY{Milton:1995:WET}  were recently experimentally realized by \citeAPY{Kadic:2012:PPM} dramatically 
illustrating the present day ability to tailor make designer microstructures (see also  \citeAPY{Buckmann:2012:TMM}).

The concepts of  unimode, bimode, trimode, quadramode and pentamode materials can be extended to non-linear elasticity and find a natural place
when one is interested in trying to find what metamaterials can be produced with  two and three dimensional periodic arrays of rigid bars and pivots.
Since they are periodic they have an  underlying Bravais lattice.  A two dimensional Bravais lattice consists of points 
\beq \Bx=i\Bu+j\Bv, \eeq{1.1}
as $i$ and $j$ range over all integers. Let $\BF$ be the deformation matrix with the primitive vectors $\Bu$ and $\Bv$ as columns. Then under a rotation $\BR$, since $\Bu$ and $\Bv$ transform
to $\BR\Bu$ and $\BR\Bv$ it follows that $\BF$ transforms to $\BR\BF$ leaving the symmetric Cauchy-Green matrix $\BA=\BF^T\BF$ invariant. There are
many other Bravais lattices. In particular, there are lattices with lattice vectors
\beq \Bu'=k\Bu+\ell\Bv, \quad \Bv'=m\Bu+n\Bv, \eeq{1.1a}
for any choice of the integers $k$, $\ell$, $m$ and $n$ such that $\Bu'$ and $\Bv'$ are independent. 
The corresponding deformation matrix $\BF'$, which has $\Bu'$ and $\Bv'$ as columns, is given by
\beq \BF'=\BF\BM,\quad {\rm where}~
\BM=\begin{pmatrix} k & m \\ \ell & n \end{pmatrix}, 
\eeq{1.1b}
and $\BM$ is non-singular. Other Bravais lattices will correspond to matrices $\BM$ with elements that are not necessarily integers.
As the material continuously deforms, with the Bravais lattice having lattice vectors $\Bu$ and $\Bv$ 
undergoing an affine transformation, $\BF$ follows some trajectory $\BF(t_0)$ beginning at $t_0=t_0^-$ and ending at $t_0=t_0^+$. Associated with the deformation
is a path
\beq \BC(t_0)=[\BF(t_0^-)^T]^{-1}\BF(t_0)^T\BF(t_0)[\BF(t_0^-)]^{-1} \eeq{1.1c}
in the three dimensional space of symmetric matrices beginning at $\BC(t_0^-)=\BI$.
If we chose a different Bravais, such as that in \eq{1.1a}, this path, for the same deformation, would be the same since
\beq [\BM^T\BF(t_0^-)^T]^{-1}\BM^T\BF(t_0)^T\BF(t_0)\BM[\BF(t_0^-)\BM]^{-1}=[\BF(t_0^-)^T]^{-1}\BF(t_0)^T\BF(t_0)[\BF(t_0^-)]^{-1}. \eeq{1.1d}
On the other hand there can be continuous deformations in which the larger lattice undergoes an affine transformation, but the
smaller lattice is distorted. So it is possible that there exist additional paths of deformation $\BC(t_0)$ for the larger lattice which
are not accessible to the smaller lattice, see figure \fig{00} and the examples in  \citeAPY{Kapko:2009:CLI}, for instance. If there exist deformations 
which are non-affine at the macroscopic scale, then the structure is said to be non-affine (see figure \fig{00}(c)). In this paper we restrict our attention to affine
materials, i.e. materials for which the only macroscopic deformations are affine ones. More precisely a material is affine if and only if any deformation $\Bx'(\Bx)$ (defined only for points
$\Bx$ on the rigid bars which get moved to $\Bx'(\Bx)$) necessarily approaches $\BB\Bx+\Bb$ as $|\Bx|\to\infty$, for some non-singular matrix $\BB$ and vector $\Bb$ (dependent on the deformation) 
to within terms of $o(|\Bx|)$. A material which isn't affine is non-affine. 

\begin{figure}[htbp]
\vspace{1.5in}
\hspace{0.0in}
{\resizebox{1.0in}{0.5in}
{\includegraphics[0in,0in][6in,3in]{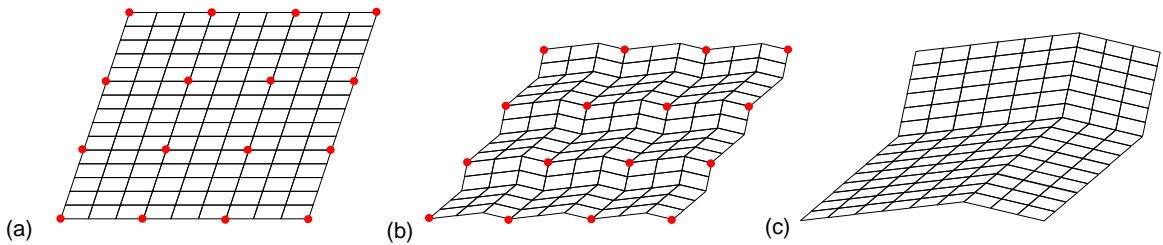}}}
\vspace{0.1in}
\caption{A parallelogram array of rigid bars can deform from (a) to (b). The red dots denote the Bravais lattice which undergoes
an affine transformation. As a result the array is not unimode, but instead trimode. It is a non-affine material as it has 
macroscopic deformations which are non-affine as in (c).}
\labfig{00}
\end{figure}

A material is classed as null-mode if the
only possible continuous deformation path for any Bravais lattice is the point $\BC(t_0)=\BI$. Some internal motions could be possible but
macroscopically the material is rigid to periodic deformations (in the sense that for any Bravais lattice the only continuous deformations that keep
the periodicity of that lattice only rotate and translate the lattice). A material is classed as unimode if it is not null-mode and the possible deformations
$\BC(t_0)$ for any Bravais lattice all lie on the same one dimensional curve. A material is classed as bimode if it is not null-mode or unimode 
and the possible deformations $\BC(t_0)$ for any Bravais lattice all lie on the same two dimensional surface. 
Four examples of such two-dimensional unimode materials that may be constructed from bars and pivots are given in figure \fig{0}, with underlying Bravais lattices marked by the red dots. The kagome lattice also appears to be a unimode material for which the only macroscopic mode of deformation is a dilation [see, for example, \citeAY{Grima:2006:ABR}; \citeAY{Hutchinson:2006:SPP}; \citeAY{Kapko:2009:CLI}; and \citeAY{Sun:2012:IAR}].

\citeAPY{Guest:2003:DRS} investigated the question as to whether it is possible to build a 
periodic pin-jointed structure having no easy modes of deformation and such that replacing any bar by an actuator and changing its length leads only to a change of the geometry of the structure, and not to self stress. 
They pointed out this is of interest to designing adaptive structures. They found that no such periodic rigid pin-jointed structures can exist.
However the periodic unimode materials considered here (unlike the kagome lattice)
have the property that they become macroscopically rigid if a single (appropriately placed) bar 
is added to the structure. Changing 
the length of this bar (within limits) leads only to a change of the geometry of the structure, and not to self stress. Replacing that bar by an actuator
leads to interesting adaptive structures. 

A three dimensional Bravais lattice consists of points 
\beq \Bx=i\Bu+j\Bv+k\Bw, \eeq{1.2}
as $i$, $j$ and $k$ range over all integers, and we take $\BF$ as the matrix with the primitive vectors $\Bu$, $\Bv$ and $\Bw$ as columns. Then as the Bravais lattice
undergoes a continuous affine transformation, the matrix $\BF$ varies along a path $\BF(t_0)$ beginning at $t_0=t_0^-$ and ending at $t_0=t_0^+$, and the symmetric 
matrix $\BC(t_0)$ defined by \eq{1.1c} follows a trajectory in six dimensional space. 
In three dimensional unimode, bimode, trimode, quadramode and pentamode materials the possible trajectories
$\BC(\Bt_0)$ are confined to, respectively, a one, two, three, four, or five dimensional manifold, for all Bravais lattices and
for all possible affine deformations of those lattices.  

Here we only consider unimode materials built from rigid bars and pivots and address the problem of characterizing what trajectories $\BC(t_0)$ are realizable and in the process
we find multiscale microstructures which realize given trajectories $\BC(t_0)$ to an arbitrarily high degree of approximation. We do not consider trajectories
where the microstructure degenerates into a lower dimensional geometry, and thus we assume $\BC(t_0)$ remains strictly positive definite,
and so the trajectory $\BC(t_0)$ is confined to lie within the ``cone'' of positive definite matrices. Here we essentially show, by construction, that every such trajectory is 
realizable to an arbitrarily high degree of approximation: there are no hidden constraints. 

Most of the structures we consider here are currently impractical to build because of their highly multiscale nature and because we assume the bars are perfectly rigid, and the junctions pin-like. Nevertheless, with the current rate of progress in rapid prototyping of complex structures, as exemplified by \citeAPY{Kadic:2012:PPM},
it is hard to predict how well they might be approximated in the future. Also 
 they show what is theoretically possible, and this should motivate the design of more realistic materials exhibiting a wide range of interesting behaviors. 

In the course of the analysis we construct two and three dimensional dilational materials with an arbitrarily large flexibility window as defined by \citeAPY{Sartbaeva:2006:FWZ}.
Dilational materials are perfect auxetic materials for which a dilation is the only easy mode of deformation: 
see the two-dimensional examples of figures \fig{0}(b) and \fig{0}(c),
and, in three dimensions, those of \citeAPY{Buckmann:2012:RTD}. In my dilational material with an arbitrarily large flexibility window the cell size could (theoretically)
expand from the size of a pea to the size of a house!

Historically the study of what mechanisms can be achieved by linkages of rigid bars and pivots 
attracted considerable attention. In particular it was shown 
(\citeAY{Kempe:1875:GMD}; \citeAY{Artobolevskii:1964:MGP}; \citeAY{Kapovich:2002:UTC})
that any planar algebraic curve could be drawn by a linkage of bars and pivots (allowing for
the bars to slide over one another so that the structure is not strictly planar). By contrast 
our focus is on the (macroscopic) mechanisms of periodic structures of bars and pivots.

\begin{figure}[htbp]
\vspace{4.0in}
\hspace{0.5in}
{\resizebox{1.0in}{0.5in}
{\includegraphics[0in,0in][7in,3.5in]{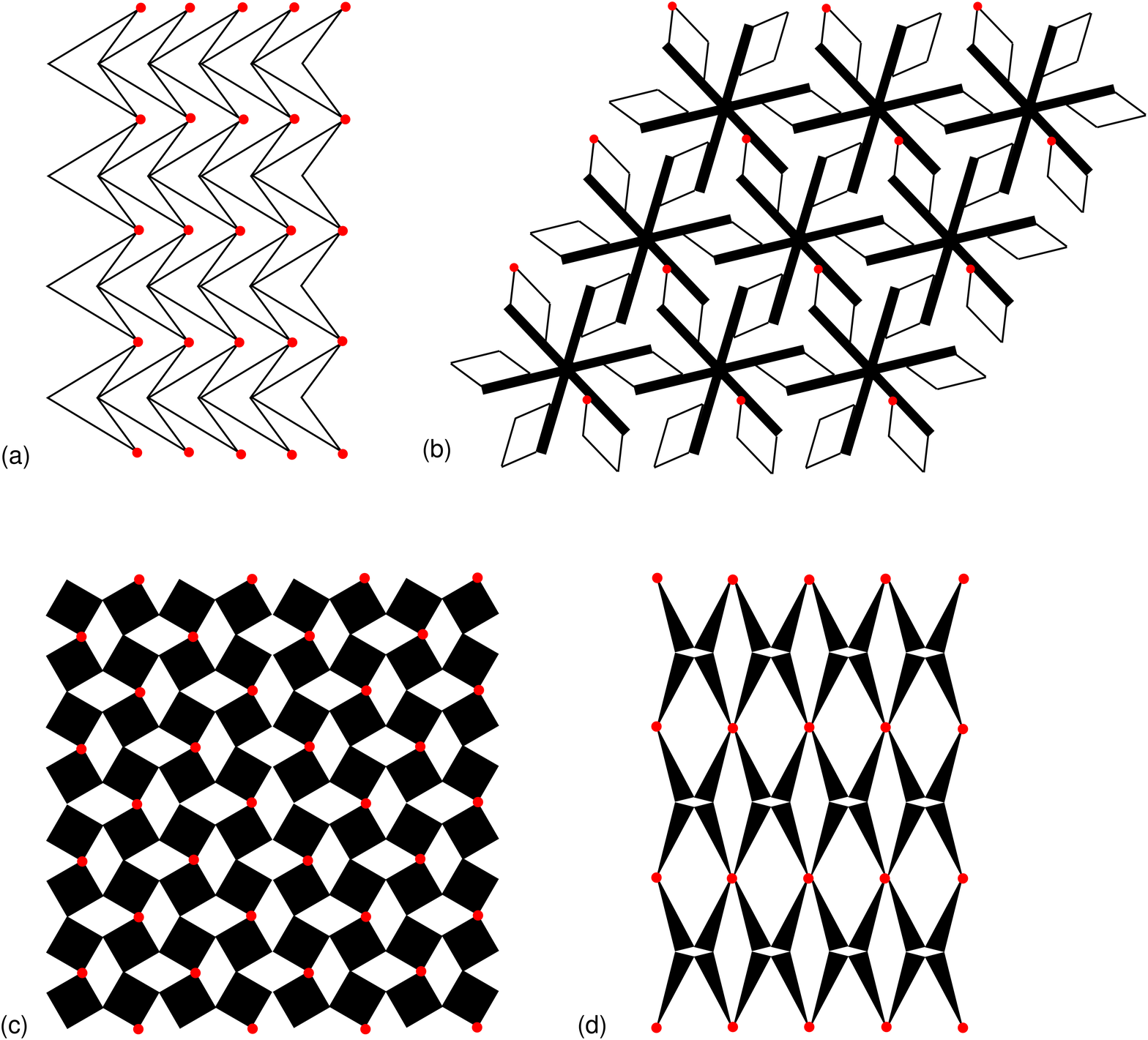}}}
\vspace{0.1in}
\caption{Four examples of unimode materials. The example (a) of {\protect\citeAPY{Larsen:1997:DFC}}, which is a simplified version of a structure of {\protect\citeAPY{Milton:1992:CMP}},
has an incremental Poisson ratio which is negative in the configuration shown. The examples (b) of {\protect\citeAPY{Milton:1992:CMP}} and (c) of
{\protect\citeAPY{Grima:2000:ABR}} are dilators, for which the only mode of macroscopic deformation is a dilation. Example (d) is an expander
which will be analyzed in detail in the next section. In (b), (c) and (d) the black regions are rigid polygons, which may be replaced by trusses of bars 
as shown in figure \fig{0a}. The vertices in (a) where lines (rigid bars) meet, or in (b) where a line (rigid bar) meets a polygon, or in (c) and (d) where 
two or more polygons meet, are hinge joints (pivots). In each example an underlying Bravais lattice is marked by the vertices with red dots.}
\labfig{0}
\end{figure}

\begin{figure}[htbp]
\vspace{1.5in}
\hspace{1.0in}
{\resizebox{2.0in}{1.0in}
{\includegraphics[0in,0in][8in,4in]{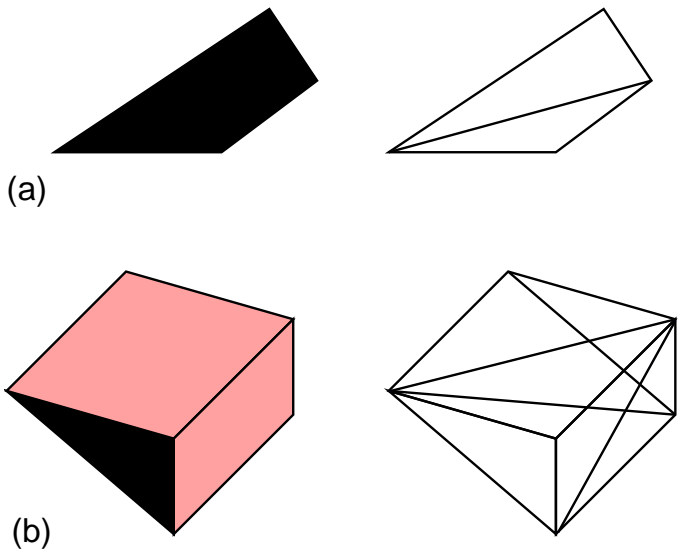}}}
\vspace{0.1in}
\caption{As is well known a solid rigid polygon in two dimensions can be replaced by a truss of bars, as shown in (a) for a quadrilateral. Similarly  a solid rigid polyhedron in three dimensions
can be replaced by a truss of bars, as shown in (b) for a triangular prism: the basic idea is to subdivide the polyhedron into tetrahedra, and then replace the tetrahedra by bars along
their edges. To simplify subsequent figures we will use rigid polygons (colored black) and rigid polyhedra (colored  black
and pink) rather than the equivalent truss.} 
\labfig{0a}
\end{figure}

\section{Realization of planar rectangular materials with arbitrary response}
\setcounter{equation}{0} 

In this section we only consider planar materials for which $\Bu$ and $\Bv$ are orthogonal and remain so as the material deforms. We call these rectangular materials even though the interior microstructure
might not necessarily have rectangular symmetry. We let $\Gl_1=|\Bu|$ and $\Gl_2=|\Bv|$. So for such materials the question becomes: what trajectories $(\Gl_1(t_0),\Gl_2(t_0))$
can be realized, as $t_0$ varies in the interval $t_0^-\leq t_0\leq t_0^+$. Such trajectories must lie in the quadrant $\Gl_1>0$ and $\Gl_2>0$. We will show that
every such trajectory can be approximately realized. Since we want to attach together unit cells which have different microstructures it is simplest to require that the parallelogram unit
cell (which when periodically repeated forms the material) has all its microstructure remain inside the unit cell as the cell deforms
and only touches the boundary of the unit cell at the vertices of the parallelogram which we call the support points. 
This is true for the microstructure of figure \fig{0}(d) but not for the microstructures of
figures \fig{0}(a), \fig{0}(b), and \fig{0}(c). This is not a restrictive assumption as for most microstructures, such as those in figures \fig{0}(a), \fig{0}(b) and \fig{0}(c), we can take
a large (finite) array of cells and attach bipod supports to the corners of the array and  rescale the microstructure to obtain a new cell with almost the same properties
as the original cell but for which the microstructure does not extend outside the unit cell: see figure \fig{1} to get the idea.

\subsection{Some elementary operations on realizable functions}

To begin with we mainly concentrate on  trajectories $(\Gl_1(t_0),\Gl_2(t_0))$ for which $\Gl_2$ is a single valued continuous function $\Gl_2=f(\Gl_1)$ of $\Gl_1$.
As we will see there are many operations that can be done with realizable functions to obtain new realizable functions. We will not cover all of them, only
those necessary to establish the main result of the paper, that all trajectories are approximately realizable. We start with the elementary (and obvious) ones,

\begin{itemize}

\item {\bf Rescaling}: If $\Gl_2=f(\Gl_1)$ is realizable then so is $\Gl_2=\Ga^{-1}f(\Ga\Gl_1)$, for any real constant $\Ga>0$, by rescaling the microstructure in the cell.

\item {\bf Rotation}: If $\Gl_2=f(\Gl_1)$ is realizable then so is $\Gl_1=f(\Gl_2)$ by rotating the cell (and its microstructure) by $90^\circ$. This is most useful when
$\Gl_2$ is a vertical distance which determines the horizontal distance $\Gl_1$ though the function $f$.

\item {\bf Reshaping}:  If $\Gl_2=f(\Gl_1)$ is realizable and $f$ is continuous, $\Gl_2=rf(\Gl_1)$ is realizable for any rational $r=p/q$, where
$p$ and $q$ are integers. To see this, for simplicity in the case where $p=3$ and $q=2$, consider an array of $np$
cells in the vertical direction and $nq$ cells in the horizontal direction, as illustrated in figure \fig{2} with $n=4$, where the
bipod supports have leg length $d$ (and have a pivot joint at the midpoint). By rescaling the microstructure of the material we can assume $g=(nq)^{-1}f(nqh)$.
Also let the bipod supports scale as  $d=d_0/\sqrt{n}$, so they are small compared with $\Gl_1$ and $\Gl_2$ but large compared with
the cell size. Then as $n\to\infty$ we have $nqh\to \Gl_1$ and $npg=r f(nqh) \to \Gl_2$. So in the limit we see that $\Gl_2=rf(\Gl_1)$.

\item {\bf Addition}: If $\Gl_2=f_1(\Gl_1)$ and  $\Gl_2=f_1(\Gl_1)$ are both realizable and continuous, then  
$\Gl_2=f_1(\Gl_1)+f_2(\Gl_1)$ is realizable. Referring to figure \fig{3}, illustrated for the case $n=8$, by rescaling the microstructure of the material we can assume $g_1=(n)^{-1}f_1(nh)$ and 
$g_2=(n)^{-1}f_2(nh)$. We again let the supports scale as  $d=d_0/\sqrt{n}$.  Then as $n\to\infty$ we have $nh\to \Gl_1$ and $ng_1+ng_2=f_1(nh)+f_2(nh)\to\Gl_2$.
In the limit we obtain $\Gl_2=f_1(\Gl_1)+f_2(\Gl_1)$.

\end{itemize}
\begin{figure}[htbp]
\vspace{2.5in}
\hspace{1.0in}
{\resizebox{2.0in}{1.0in}
{\includegraphics[0in,0in][6in,3.0in]{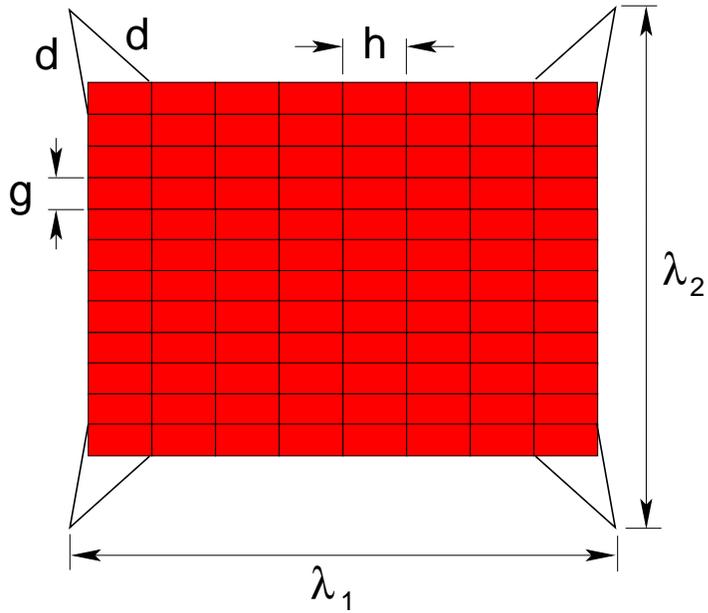}}}
\vspace{0.1in}
\caption{The configuation for reshaping a cell, showing that for any rational $r=p/q$,  $\Gl_2=(p/q)f(\Gl_1)$ is realizable
if  $\Gl_2=f(\Gl_1)$ is realizable. The configuration is illustrated for the case $n=4$, $p=3$ and $q=2$, and one should
take the limit $n\to \infty$. The leg length $d$ of the bipod corner supports has to be scaled appropriately as discussed 
in the text.}
\labfig{2}
\end{figure}

\begin{figure}[htbp]
\vspace{3.0in}
\hspace{1.0in}
{\resizebox{2.5in}{1.25in}
{\includegraphics[0in,0in][6in,3.0in]{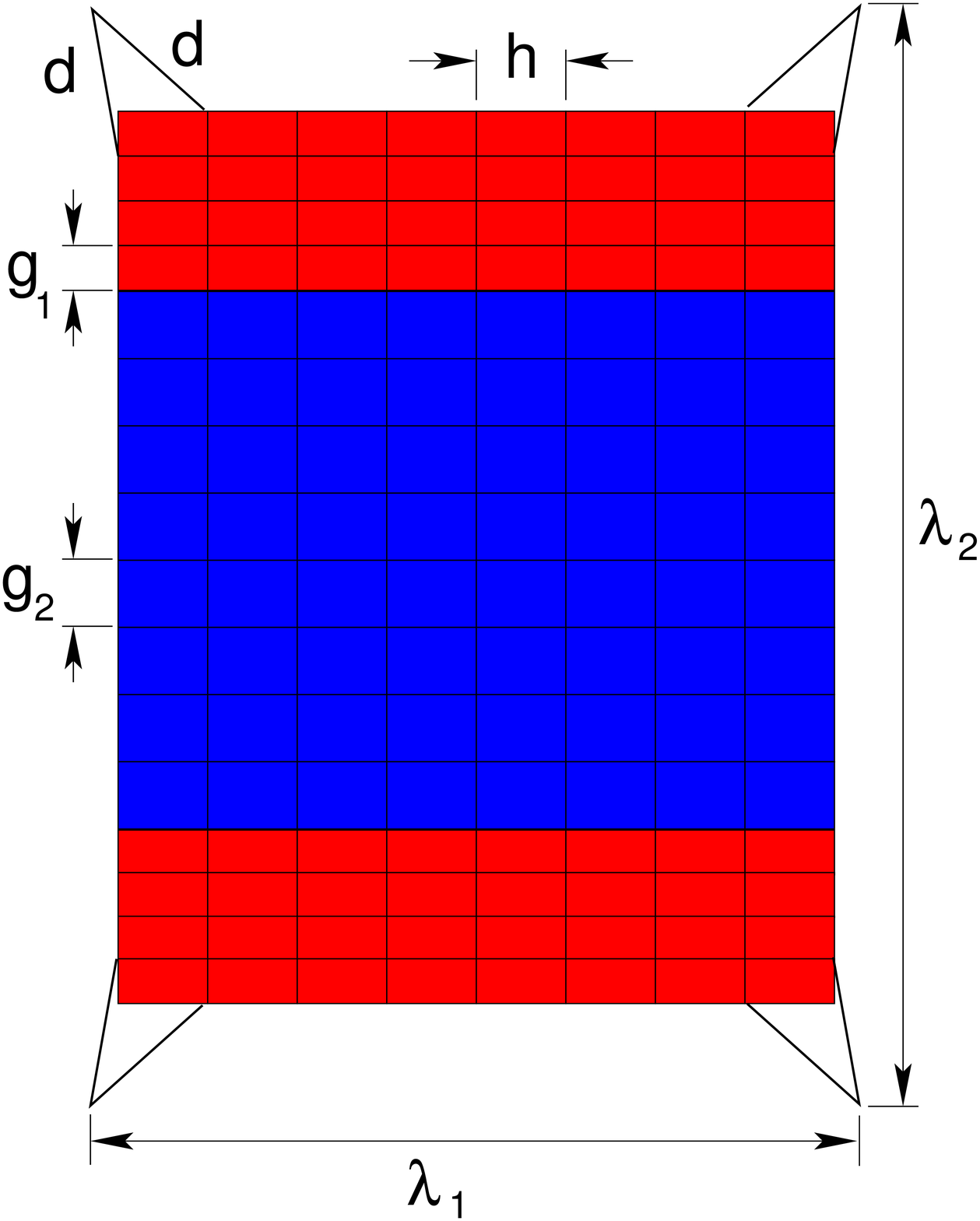}}}
\vspace{0.1in}
\caption{The configuration for an adder, combining two materials (red and blue) to obtain a metamaterial with a response function which is the sum of the
response functions of the two materials. The bipod corner supports have leg length $d$ which is large compared with $h$, $g_1$ and $g_2$  yet small compared with $\Gl_1$ and $\Gl_2$. }
\labfig{3}
\end{figure}

\subsection{Some elementary realizable functions}

We obviously need some elementary realizable functions to begin building more complicated ones. 
Let us examine in more detail the response of the expander of figure \fig{0}(c). The unit cell is shown in more detail
in figure \fig{1}(a), which defines various parameters. Elementary trigonometry shows that
\beq \Gl_1=2a\sin\Ga+2\Gve\cos\Ga,\quad \Gl_2=2a\cos\Ga,\eeq{2.1}
implying that $(\Gl_1,\Gl_2)$ lies on the portion of the ellipse
\beq (a\Gl_1-\Gve\Gl_2)^2=a^2(4a^2-\Gl_2^2), \eeq{2.2}
with $2\sqrt{\Gve^2+a^2}>\Gl_1> 2\Gve$ and $2a>\Gl_2>2a\Gve/\sqrt{\Gve^2+a^2}$ where the last constraint ensures that the triangles remain in the cell. 

\begin{figure}[htbp]
\vspace{4.5in}
\hspace{1.0in}
{\resizebox{2.0in}{1.0in}
{\includegraphics[0in,0in][6in,3in]{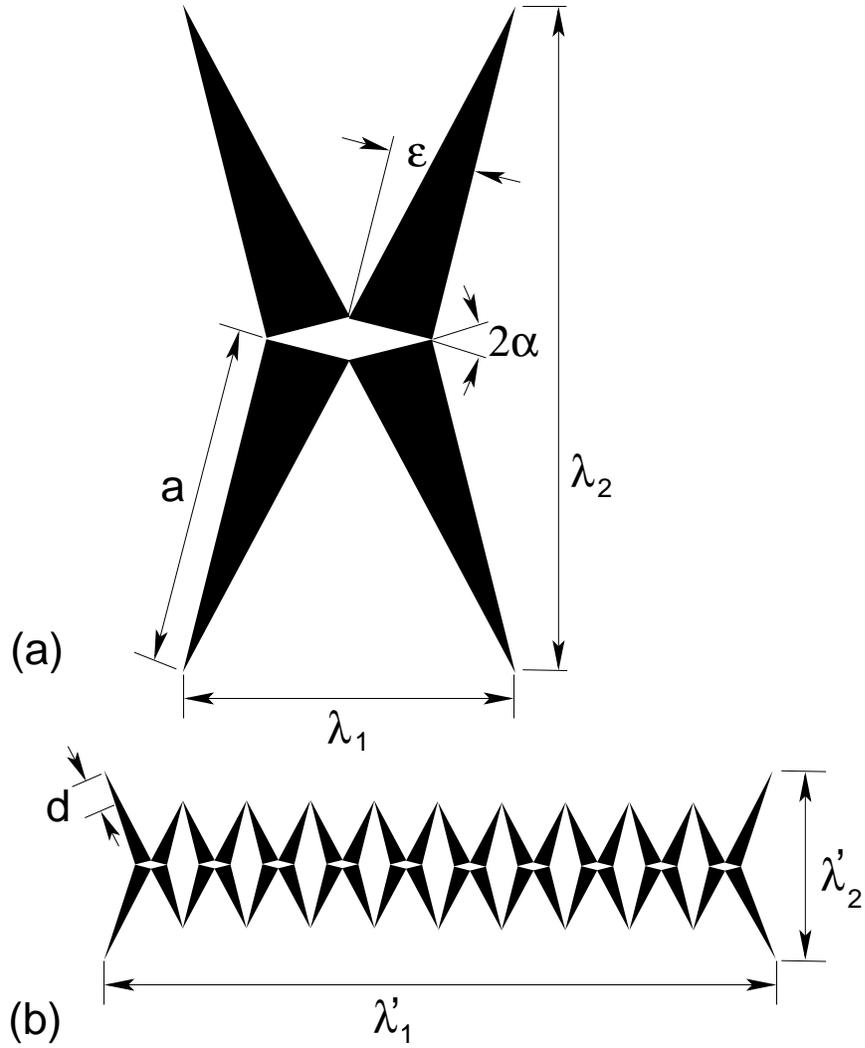}}}
\vspace{0.1in}
\caption{Shown in (a) is the basic geometry of a cell of an expander. Here $\Gve$ should be chosen very small. The black right angled
triangles are rigid. Shown in (b) is an expander of $n=10$ cells, with the legs of the corner triangles chosen slightly longer, so they
can be used as supports.}
\labfig{1}
\end{figure}

In particular let us take $n$ of these elements, arranged as in figure \fig{1}(b), to form a new cell with dimensions $(\Gl_1',\Gl_2')$ and let us set $\Gve=\Gve_0/n^2$ and $d=d_0/n^2$
keeping $a$ independent of $n$. Then given a domain $D$ of operation, $\Gl_1^-<\Gl_1'<\Gl_1^+$, we choose $n$ sufficiently large that the minimal extension of the expander is less
than $\Gl_1^-$ and its maximum extension is greater than $\Gl_1^+$. Then for any given $\Gl_1'$ in $D$ we have 
\beq 0<(a\Gl_1-\Gve\Gl_2)<a\Gl_1<a\Gl_1'/n<a\Gl_1^+/n,\quad \Gl_2'<\Gl_2<\Gl_2'+d_0/n^2. \eeq{2.3}
These estimates in conjunction with \eq{2.2} imply that for any $\Gl_1'$ in $D$, $\Gl_2'$ converges to $2a$ as $n\to\infty$, independent of $\Gl_1'$.
In other words any positive constant function is (approximately) realizable on a given domain $D$. 
Such a cell behaves like an ideal expander: when $n$ is very large we can adjust the horizontal length as we please (within the domain $D$) 
and the vertical length remains (essentially) unchanged. From a practical viewpoint we caution that the structure will likely be very sensitive
to small imperfections when $n$ is large. 
 
Another important cell is a dilator. An example of one, based on the chiral geometries of \citeAPY{Milton:1992:CMP}, \citeAPY{Prall:1996:PCH}, and \citeAPY{Mitschke:2011:FAF},
is shown in figure \fig{4} in four states of deformation. If the innermost square has side length
$\ell$ and the right angled triangle has height $h$ and base $b<\ell$, then the cell dimensions are 
\beq \Gl_1=\Gl_2=t_0\equiv\sqrt{2h^2+\ell^2-2\sqrt{2}h\ell\cos(\Gt+\pi/4)}, \eeq{2.4}
having a minimum value of 
\beq t_0^-=\sqrt{2h^2+\ell^2-2h\ell} \eeq{2.4a}
at $\Gt=0$ and a maximum value of 
\beq t_0^+=\ell+\sqrt{2}h  \eeq{2.4b}
at $\Gt=3\pi/4$ corresponding to case $(b)$ in figure  \fig{4}. We call $(t_0^-,t_0^+)$ the performance range of the dilator. Rescaling the microstructure allows 
us to get performance ranges $(kt_0^-,kt_0^+)$ for any constant $k>0$. 

Now by combining four expanders
with this basic dilator cell, as sketched in figure \fig{5},  one can obtain dilators having an arbitrarily large performance ratio $t_0^+/t_0^-$
of maximum extension to minimum extension. To see this, suppose each expander has $n$ unit cells and let the 
dimensions of the inner basic dilator scale as $\ell=\ell_0/n$, $h=h_0/n$ and $b=b_0/n$, let the dimensions of the expander
scale as $a=a_0/n$, $\Gve=\Gve_0/n^2$, and let the dimensions of the bipod supports scale as $d=d_0/n$. Choose $a_0$ in the range
\beq \sqrt{2h_0^2+\ell_0^2-2h_0\ell_0}/\sqrt{2}<a_0<(\ell_0+\sqrt{2}h_0)/2, \eeq{2.5}
so that the inner basic dilator can never reach its maximum extension, but the expander can attain its minimal extension at $\Ga=0$ and
can have an angle $\Ga$ at least as large as $45^\circ$. (This inequality implies $\ell_0$ and $h_0$ must be chosen with $6h_0\ell_0>2h_0^2+\ell_0^2$ as
may be satisfied with $h_0=\ell_0$, for example).
 Choose $d_0>\sqrt{2} a_0$ so that the microstructure is always contained in the cell. 
When the expanders are fully compressed they have a length of $2n\Gve=2\Gve_0/n$, and since the maximum possible dimensions of the inner basic dilator scale as 
$1/n$ and the supports scale as $1/n$ it follows that the minimum extension scales as $1/n$. By contrast with $\Ga=45^\circ$ the expanders have an 
extension of $\sqrt{2} a_0$ in the limit $n\to \infty$. Therefore the performance ratio of maximum extension to minimum extension can be as large as one pleases with
this microstructure by choosing $n$ large enough. Thus the function $\Gl_2=\Gl_1$ is realizable, and by addition with an ideal expander so is the function $\Gl_2=\Gl_1+c$ for any
constant $c>0$.

\begin{figure}[htbp]
\vspace{4.0in}
\hspace{1.0in}
{\resizebox{3in}{1.5in}
{\includegraphics[0in,0in][6in,3.0in]{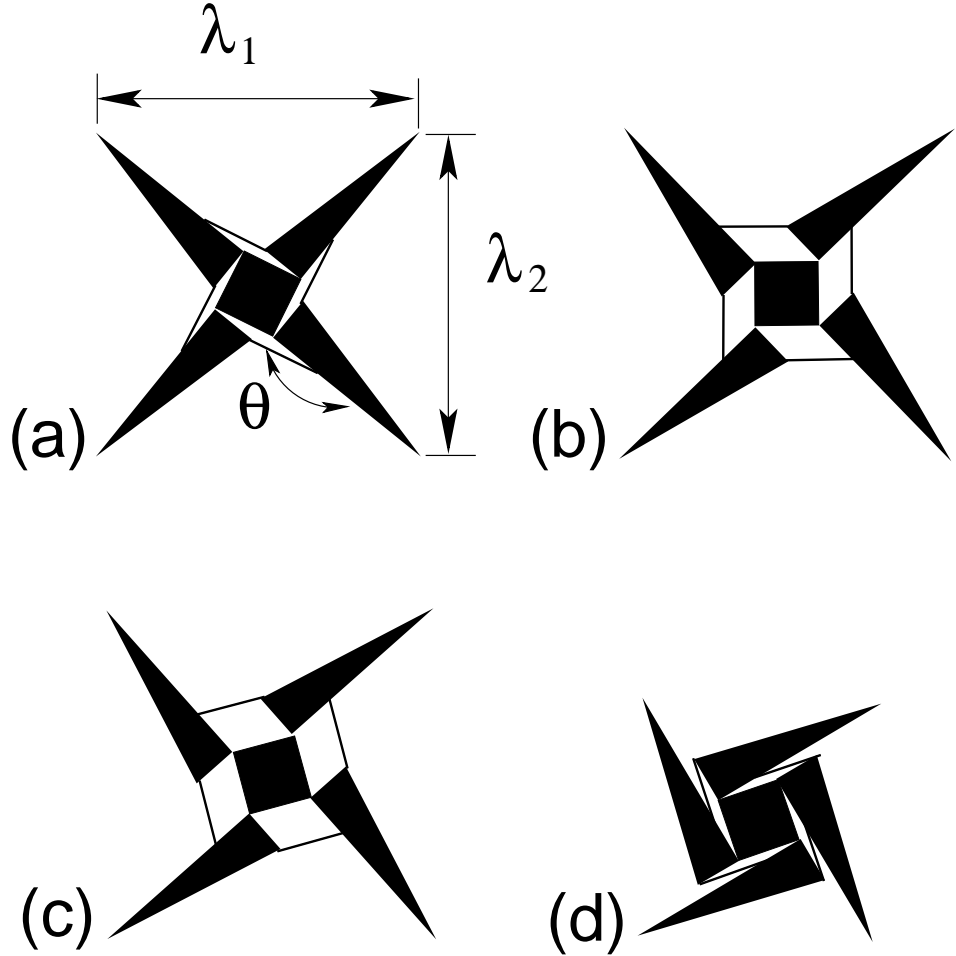}}}
\vspace{0.1in}
\caption{Four successive configurations of a basic dilator cell. The maximum area of the dilator enclosed by the square formed by the outermost points occurs in (b)
where this square is aligned with the innermost square}
\labfig{4}
\end{figure}

\begin{figure}[htbp]
\vspace{3.0in}
\hspace{1.0in}
{\resizebox{2.0in}{1.0in}
{\includegraphics[0in,0in][6in,3in]{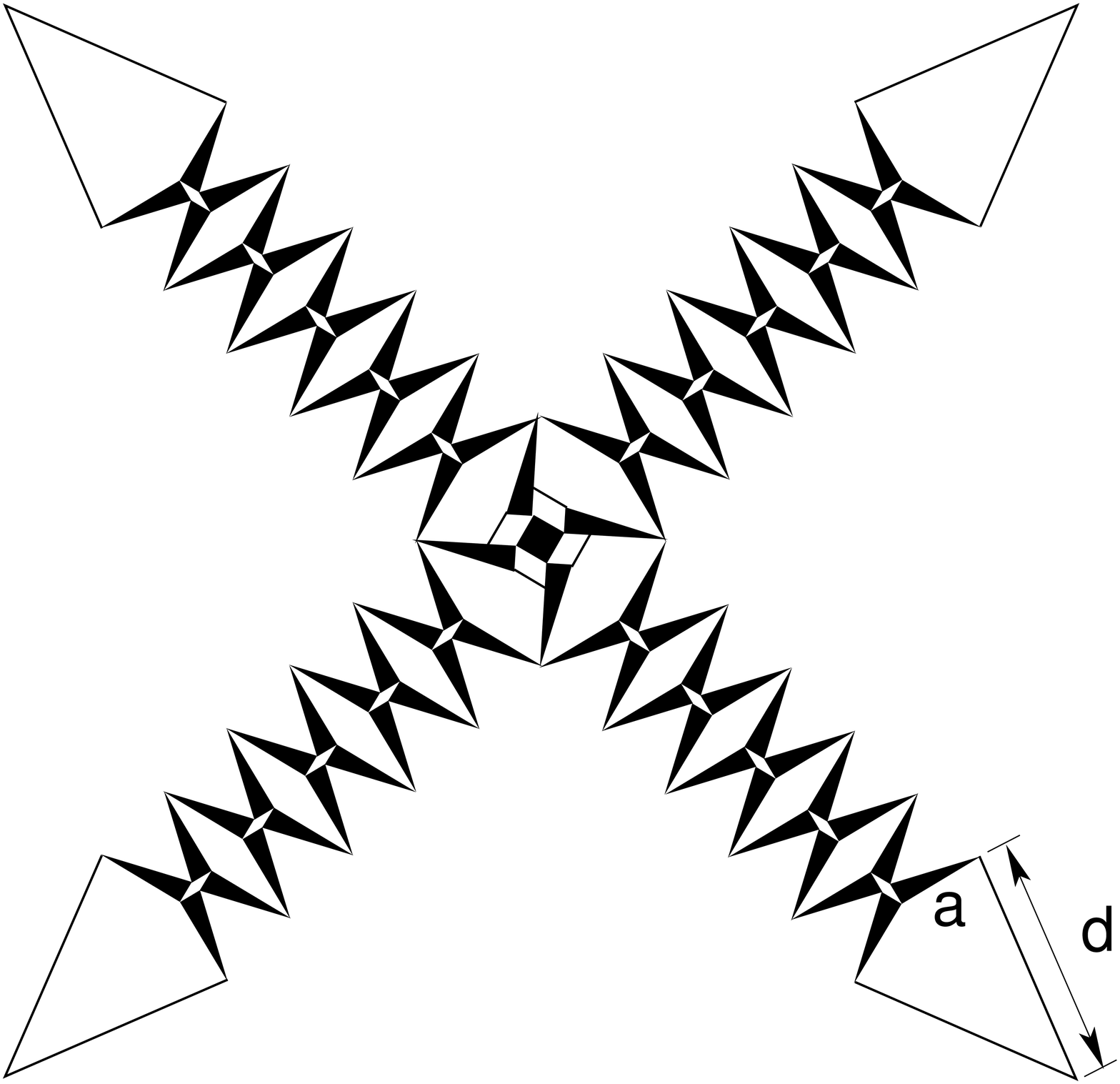}}}
\vspace{0.1in}
\caption{By combining the basic dilator cell of figure \fig{4} with four expanders one obtains dilators having an arbitrarily large ratio
of maximum extension to minimum extension.}
\labfig{5}
\end{figure}

Now that we have dilators we can build many things. 
\begin{itemize}
\item To build a material where in a domain $\Gl_2$ depends linearly on $\Gl_1$ with negative slope consider the
microstructure of figure \fig{6}. In the cell there are $n$ dilator cells in the horizontal direction and $n$ dilator cells in the vertical direction.
We only show one corner of the cell as the microstructure in the other corners are just reflections of this microstructure about the $x_1$ and $x_2$ axes. 
We scale the geometry with $a=a_0/\sqrt{n}$, $b=a_0/\sqrt{n}$, $c=c_0/\sqrt{n}$, $d=d_0/\sqrt{n}$, and $\Gf=\Gf_0/\sqrt{n}$ while varying $\Gt$ so that  
$\Gt+\Gf$ is independent of $n$. Since by trigonometry $\Ga+\Gg+\Gt=\pi$ and  $\Gt+\Gf+\Gg-\Gb=\pi$ we deduce that $\Ga+\Gb=\Gf=\Gf_0/\sqrt{n}$.
So as one might expect, the angles $\Ga$ and $\Gb$ must go to zero as $n\to\infty$. We can let $\Ga=\Ga_0/\sqrt{n}$ and $\Gb=\Gb_0/\sqrt{n}$ where
$\Ga_0>0$ and $\Gb_0>0$ depend on the deformation and $\Ga_0+\Gb_0=\Gf_0$. 
Then the length $nh$ of the dilator segment in the horizontal direction satisfies
\beq nh=2nb\sin{\Ga}=2\sqrt{n}b_0\sin{(\Ga_0/\sqrt{n})}\to 2b_0\Ga_0\quad{\rm as}~n\to\infty, \eeq{2.6}
while the length $ng$ of the dilator segment in the vertical direction satisfies
\beq ng=2nc\sin{\Gb}=2\sqrt{n}c_0\sin{(\Gb_0/\sqrt{n})}\to 2c_0\Gb_0\quad{\rm as}~n\to\infty. \eeq{2.7}
We restrict the range of operation so $\Gl_1>\Gn$ and $\Gl_2>\Gn$ for some small $\Gn>0$. Then
$(\Gl_1,\Gl_2)$ must approach $(nh,ng)$ since the corner cell geometry
scales roughly as $1/\sqrt{n}$. Finally since $\Ga_0+\Gb_0=\Gf_0$ we conclude that, in the limit $n\to\infty$,
\beq \Gl_2= 2c_0\Gf_0-(c_0/b_0)\Gl_1. \eeq{2.8}
Since the ratio $c_0/b_0$ can be any positive value we please, it is clear that we can get any negative slope. By rescaling it is clear that the
constant can take any desired positive value. 

\begin{figure}[htbp]
\vspace{4.0in}
\hspace{0.2in}
{\resizebox{2.5in}{1.25in}
{\includegraphics[0in,0in][6in,3in]{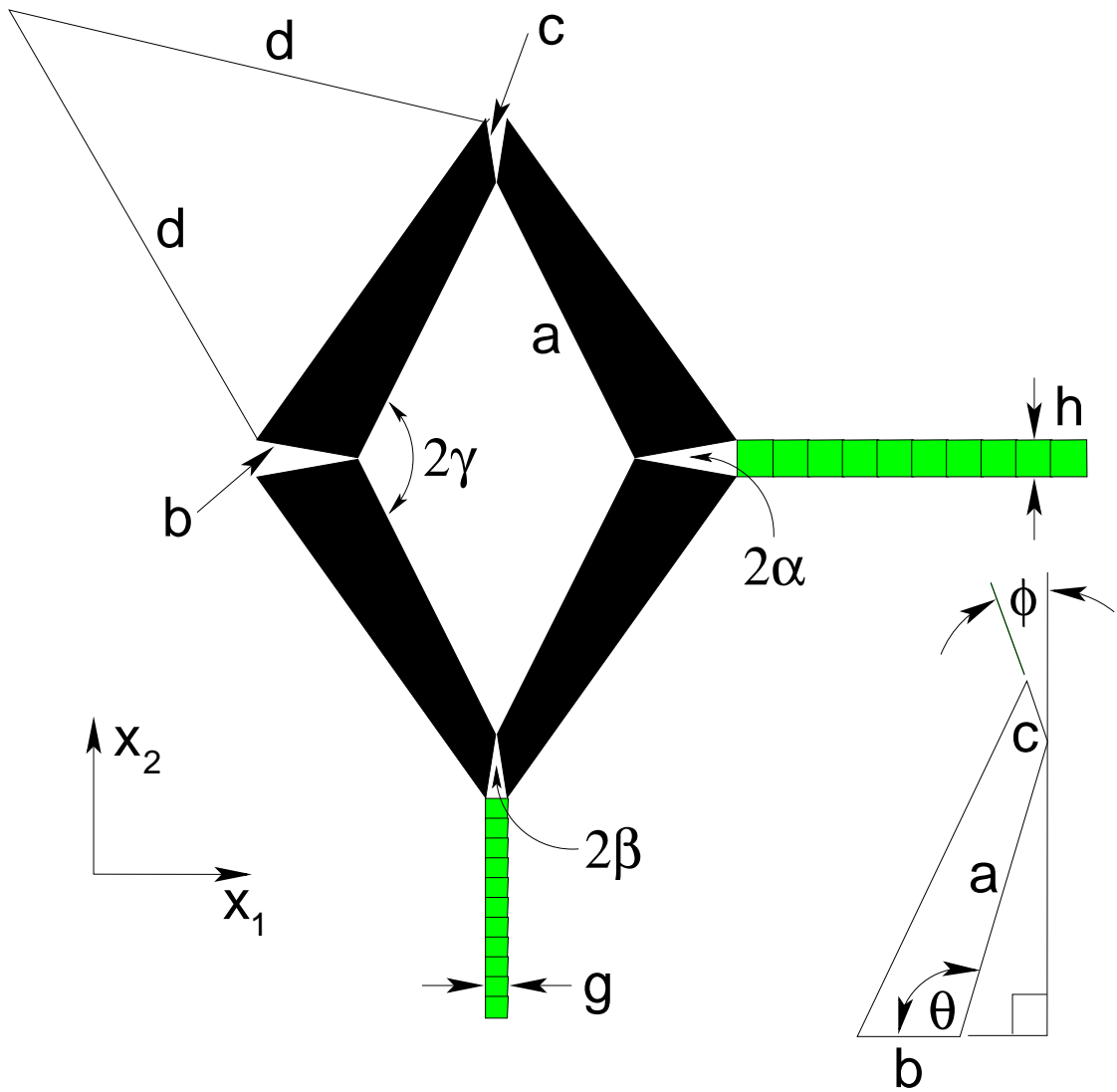}}}
\vspace{0.1in}
\caption{The corner of a cell which realizes a linear dependence $\Gl_2=r-s\Gl_1$ (where $r>0$ and $s>0$)
 to an arbitrarily high degree of approximation within a domain $\Gl_1>\Gn>0$ and $\Gl_2>\Gn>0$ . Here the green cells are dilator cells. 
The insert shows the geometry of the rigid quadrilateral, defining the angles $\Gf$ and $\Gt$}
\labfig{6}
\end{figure}

\item To build a material where in a domain $D$, $0<\Gl_1^-<\Gl_1<\Gl_1^+$, $\Gl_2$ depends quadratically on $\Gl_1$ consider the
microstructure of figure \fig{7}. In the cell there are $n^2$ dilator cells in the horizontal direction and $n^3$ dilator cells in the vertical direction.
Like in the previous example we only show one corner of the cell as the microstructure in the other corners are just reflections of this microstructure about the $x_1$ and $x_2$ axes.
Since $a$ and $b$ are sides of a triangle with opposite angles $\Gb$ and $\pi/2-\Ga$
the law of sines implies $\sin\Gb=(a/b)\cos\Ga$. To ensure that $\Gg=0$ when $\Ga=0$, in which case $\Gb=\Gt$,
we choose  $\sin\Gt=a/b$. Then we have
\beqa g & = & 2c\sin\Gg=2c\sin(\Gt-\Gb) \nonum
        & = & 2c\left[\sin\Gt\sqrt{1-(a/b)^2\cos^2\Ga}-(a/b)\cos\Gt\cos\Ga\right] \nonum
        & = & (2ca/b)\left[\sqrt{1-(a/b)^2+(h/2b)^2}-\sqrt{(1-(a/b)^2)(1-(h/2a)^2)}\right] \nonum
        & = & \frac{c h^2}{4a\sqrt{b^2-a^2}}+\mathcal{O}(h^4/a^4).
\eeqa{2.9}
Now we take a scaling with 
\beq
a=a_0/n, \quad b=b_0/n, \quad c=c_0/n,\quad \Gve=\Gve_0/n, \quad d=d_0/n,
\eeq{2.10}
while keeping $\Gt$ independent of $n$. Since the corner microstructure shrinks to zero as $n\to\infty$, given $\Gl_1$ in the domain $D$, $n^2h$ must converge to $\Gl_1$.
Hence $h/a=hn/a_0$ is of the order of $1/n$ and converges to zero as $n\to\infty$. So from \eq{2.9} we see that $n^3g$ approaches $c_0(\Gl_1)^2/[4a_0\sqrt{b_0^2-a_0^2}]$
(with the leading correction terms in \eq{2.9} contributing a term of order $1/n$), which in this limit can be identified with $\Gl_2$ because the corner microstructure shrinks to zero.
Hence we deduce that
\beq \Gl_2=k\Gl_1^2, \quad {\rm with}~k=c_0/[4a_0\sqrt{b_0^2-a_0^2}]. \eeq{2.11}
By rescaling it is clear that the constant $k$ can be chosen to have any desired value.

\begin{figure}[htbp]
\vspace{3.0in}
\hspace{0.2in}
{\resizebox{2.0in}{1.0in}
{\includegraphics[0in,0in][8in,4in]{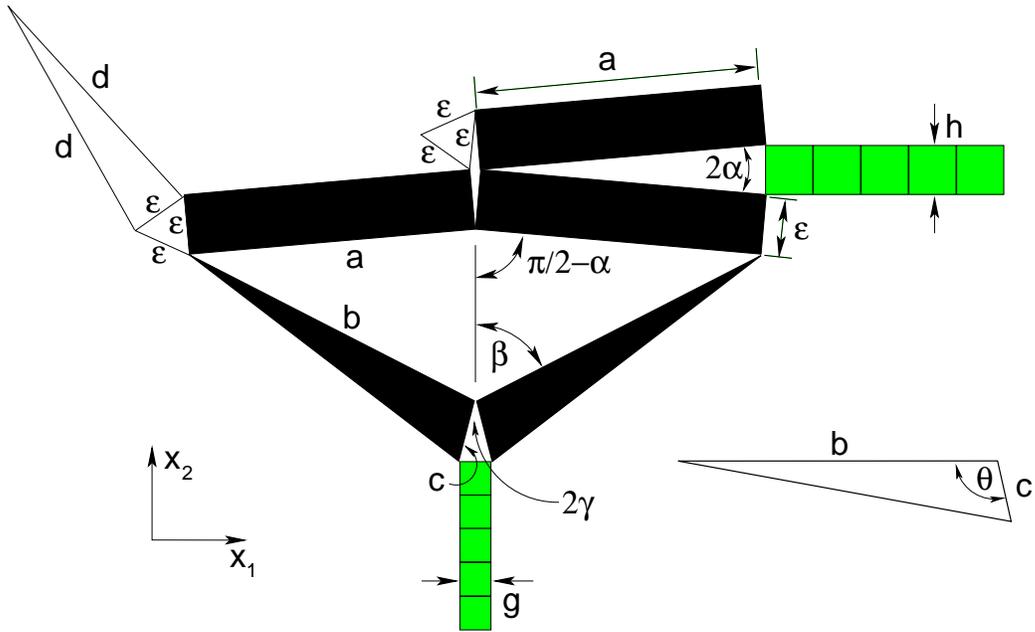}}}
\vspace{0.1in}
\caption{Design of the squarer. The corner of a cell which, within a domain, realizes a quadratic dependence $\Gl_2=k\Gl_1^2$ to an 
arbitrarily high degree of approximation within a domain. Here the green cells are dilator cells. 
The insert shows the geometry of the rigid triangles, defining the included angle $\Gt$.}
\labfig{7}
\end{figure}

\item A variant of the squarer microstructure as illustrated in figure \fig{8}, is obtained by setting $a=b$ and $\Gt=\pi/2$. We take  $n^2$ dilator cells in the horizontal direction and $n^2$ dilator cells 
in the vertical direction. In this geometry we have $h=2a\sin\Ga$ and $g=2c\sin\Ga$. This implies that for $\Gl_1$ in a domain $D$, $0<\Gl_1^-<\Gl_1<\Gl_1^+$,
in the $n\to\infty$ limit we have
\beq \Gl_2=k\Gl_1, \quad {\rm with}~k=c/a. \eeq{2.12}
Of course for rational $k$ we could have obtained the same response by reshaping. 

\begin{figure}[htbp]
\vspace{3.0in}
\hspace{0.2in}
{\resizebox{2.0in}{1.0in}
{\includegraphics[0in,0in][8in,4in]{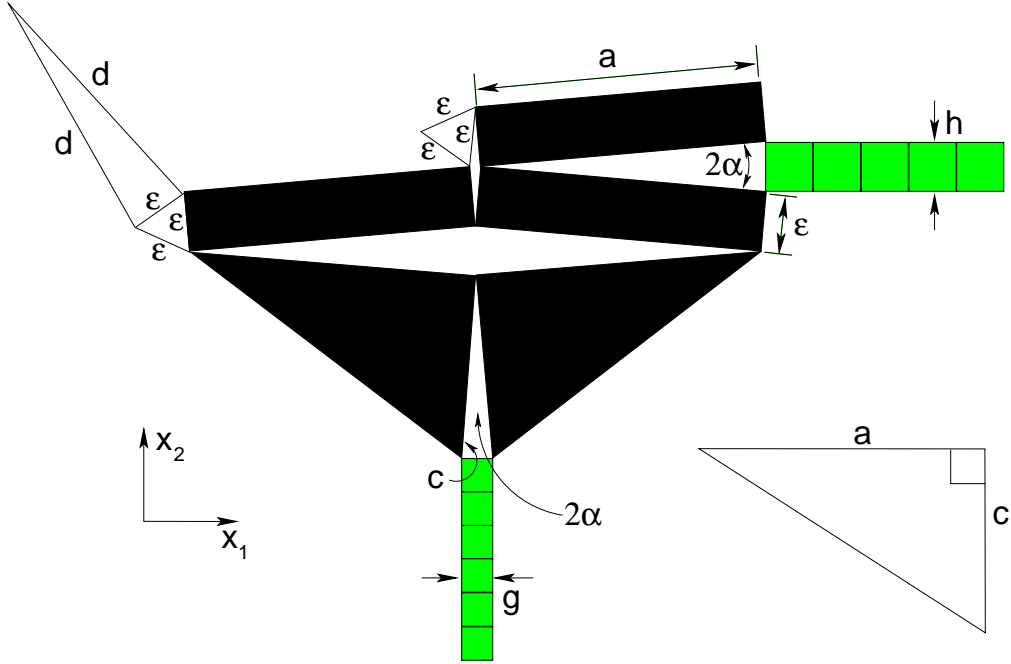}}}
\vspace{0.1in}
\caption{The corner of a cell which, within a domain, realizes a linear dependence $\Gl_2=k\Gl_1$ to an arbitrarily high degree of approximation within a domain. Here the green cells are dilator cells. 
The insert shows the geometry of the right angled rigid triangles.}
\labfig{8}
\end{figure}

\end{itemize}

\subsection{Some additional operations on realizable functions}

Let us define $\cal D$ as the set of all finite connected subintervals which are contained strictly inside the positive real axis, which may be open or closed at either end:
\beq \CD =\{ (a,b),(a,b],[a,b),[a,b]~|~0<a<b<\infty \} .\eeq{2.0}
We have the following two additional operations on realizable functions,

\begin{itemize}

\item{\bf Subtraction: } Not only can we realize the sum of two realizable continuous functions $f_1(\Gl_1)$ and $f_2(\Gl_1)$, defined (and operational) on an open domain $D\in\cal D$,
but we can also realize the difference $f_3(\Gl_1)=f_1(\Gl_1)-f_2(\Gl_1)$ on any subinterval of $D$ provided $f_3(D)\in\cal D$. We emphasize that $f_1(\Gl_1)-f_2(\Gl_1)$ could be negative outside $D$.
To see this realization consider the microstructure of figure \fig{9}.  In the cell there are $n$ dilator cells in the horizontal direction and $n$ dilator cells in the vertical direction.
We only show one corner of the cell as the microstructure in the other corners are just reflections of this microstructure about the $x_1$ and $x_2$ axes. 
Let $I=[\Gl_1^-,\Gl_1^+]$ be a closed subinterval of $D$ .  We choose to scale the 
bipod support leg length $d$ as $d=d_0/\sqrt{n}$. The inequalities 
\beq nh<\Gl_1< nh+2d= nh+2d_0/\sqrt{n} \eeq{2.12a}
imply as $n\to\infty$ that $nh$ converges to $\Gl_1$ from below. Also it implies there exists a $n_0$ (which can made explicit) such that $nh\in D$ for all $\Gl_1\in I$ when $n>n_0$.
Then $f_1(nh)$ and $f_2(nh)$ are defined. By rescaling and by a $90^\circ$ rotation of the 
microstructure of the material we can assume $s=(n)^{-1}f_1(nh)$ and $r=(n)^{-1}f_2(nh)$. By construction
$g=s-r=(n)^{-1}f_3(nh)$.
So from the inequalities
\beq ng<\Gl_2<ng+2d=ng+2d_0/\sqrt{n} \eeq{2.12b}
we see that $ng=f_3(nh)$ 
must converge to $\Gl_2$ in the limit $n\to\infty$. Therefore, since $f_3$ is continuous on $D$, in this limit $\Gl_2=f_3(\Gl_1)$ for $\Gl_1\in I$.

\begin{figure}[htbp]
\vspace{4.0in}
\hspace{0.5in}
{\resizebox{2in}{1in}
{\includegraphics[0in,0in][6in,3in]{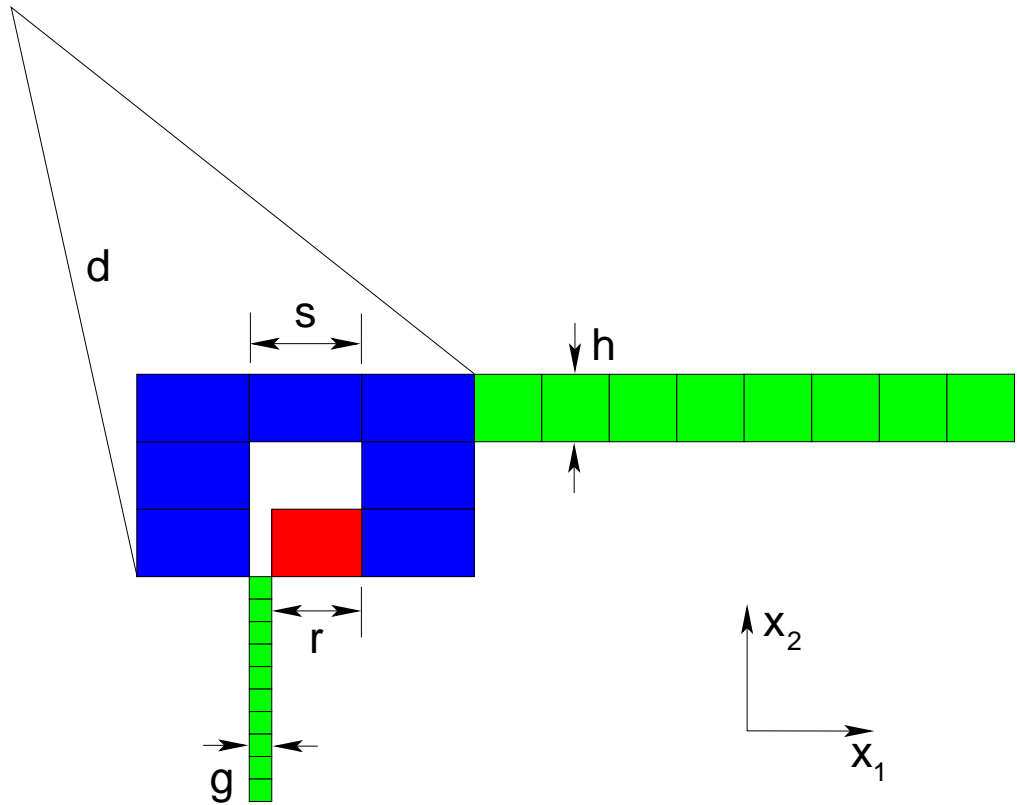}}}
\vspace{0.1in}
\caption{The configuration for the corner of a subtractor, combining two materials (blue and red) to obtain a metamaterial with a response function which is the difference of the
response functions of the two materials, within the domain of operation. Here the green cells are dilator cells.}
\labfig{9}
\end{figure}
\item{\bf Composition:} We can also compose two realizable continuous functions $f_2(\Gl_2)$ and $f_1(\Gl_1)$ to obtain a realizable function $\Gl_3=f_3(\Gl_1)\equiv f_2(f_1(\Gl_1)$.
We are given open domains $D_1\in\cal D$ and $D_2\in\cal D$, such that $f_1$ and $f_2$ are defined on $D_1$ and $D_2$, respectively,
and $f_1(D_1)\subset D_2$. Also we assume  $f_3(D_1)\in\cal D$. We consider a subinterval $I=[\Gl_1^-,\Gl_1^+]\subset D_1$.
The composition is done through the microstructure of figure \fig{10}.  In the horizontal direction there are $n$ dilator cells and in the vertical direction 
there are $n$ cells of rescaled material corresponding to $f_2$.
At the junction there is the rescaled material corresponding to $f_1$ rotated by $90^\circ$. We choose to scale the bipod support leg length $d$ as $d=d_0/\sqrt{n}$,
so the inequalities \eq{2.12a} hold and imply there exists a $n_0$  such that $nh\in D_1$ for all $\Gl_1\in I$ when $n>n_0$. Then $f_1(nh)$ and $f_3(nh)$ are defined.
By a $90^\circ$ rotation and rescaling we can realize $t=(n)^{-1}f_1(nh)$ and by rescaling realize $g=(n)^{-1}f_2(nt)=f_3(nh)$. The inequalities 
\beq ng<\Gl_2<ng+2d=ng+2d_0/\sqrt{n} \eeq{2.13}
imply that $ng=f_3(nh)$ must converge to $\Gl_3$ in the limit $n\to\infty$. Therefore, since $f_3$ is continuous on $D$, in this limit $\Gl_2=f_3(\Gl_1)$ for $\Gl_1\in I$.

\end{itemize}
\begin{figure}[htbp]
\vspace{3.0in}
\hspace{1.0in}
{\resizebox{2.5in}{1.25in}
{\includegraphics[0in,0in][6in,3in]{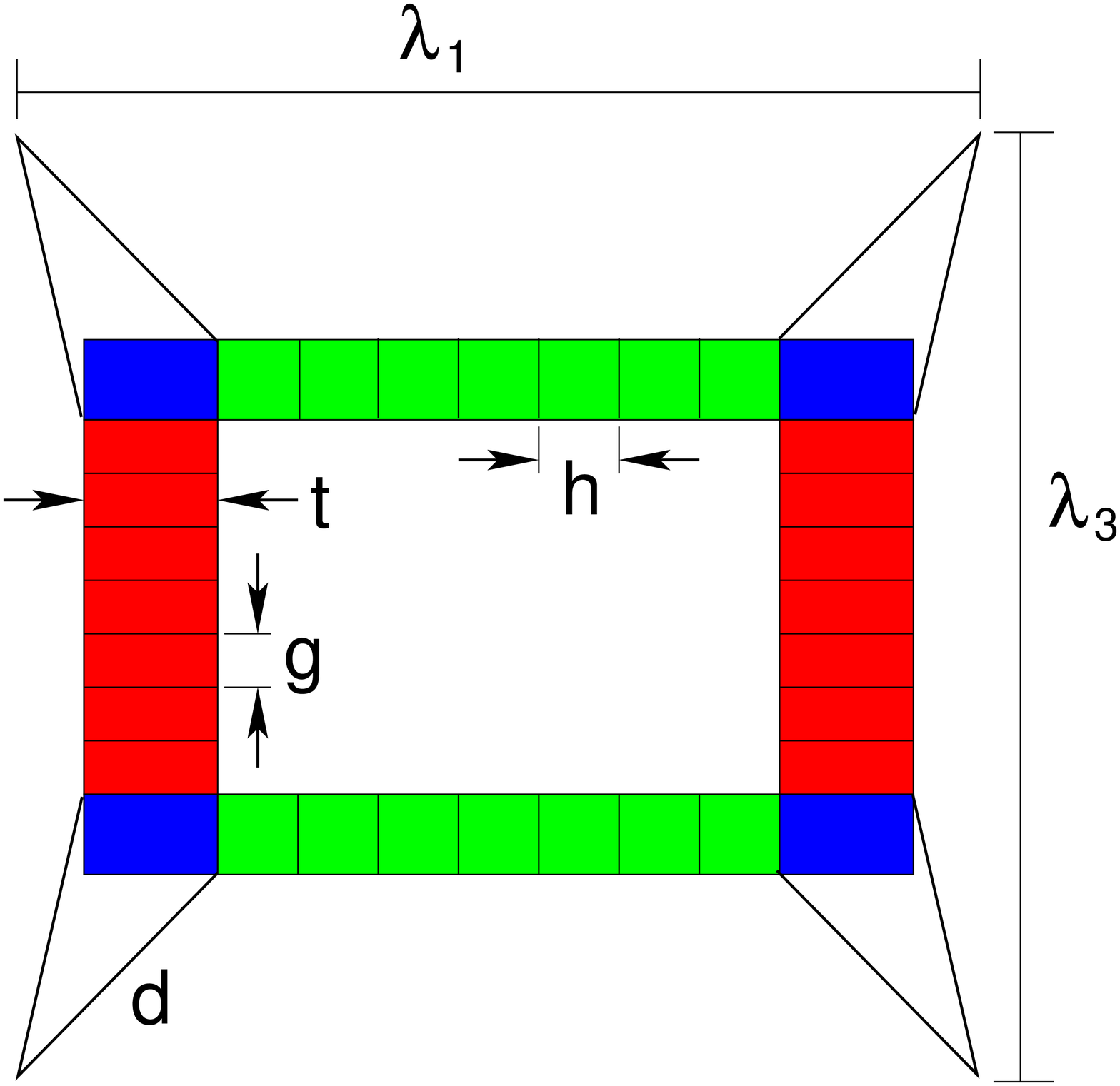}}}
\vspace{0.1in}
\caption{The configuration for a composer, combining two materials (blue and red) to obtain a metamaterial with a response function which is the composition of the
response functions of the two materials. Here the green cells are dilator cells.}
\labfig{10}
\end{figure}

\subsection{Realizability of arbitrary functions}

We first prove that one can on any subinterval $I=[\Gl_1^-,\Gl_1^+]$ of a fixed domain  $D\in\cal D$  $0<\Gl_1^-<\Gl_1<\Gl_1^+$ realize any polynomial 
\beq \Gl_2=p(\Gl_1)=a_0+a_1\Gl_1+a_2\Gl_1^2+a_3\Gl_1^3+\ldots+a_n\Gl_1^n,
\eeq{2.14}
which is bounded below by a positive constant on this domain. We prove this assertion by induction. Suppose it is true for polynomials $p$ of degree $n$ less than or equal
to $2m$, where $m\geq 0$. This is certainly true when $m=0$. Now to avoid carrying around superfluous constants let us suppose units of length (say centimeters) have been chosen,
so for example $\Gl_2=1$ means $\Gl_2$ equals one centimeter. Similarly $\Gl_2=\Gl_1^2$ means this relation holds when both $\Gl_1$ and $\Gl_2$ are measured in centimeters,
i.e. $\Gl_2=k\Gl_1^2$ where $k=1 cm^{-1}$. 
Since we can realize $\Gl_1^{m+1}$, $\Gl_1+1$, and (by the construction of the squarer of figure \fig{7})
$\Gl_1^2$ it follows by composition that we can realize (even when $m=0$)
\beq \Gl_1^{2m+2}=(\Gl_1^{m+1})^2 \quad {\rm and}~(\Gl_1+1)^{2m+2}=\Gl_1^{2m+2}+(2m+2)\Gl_1^{2m+1}+g(\Gl_1), \eeq{2.15}
where by the binomial theorem $g(\Gl_1)$ is a polynomial of degree $2m$. By rescaling we can also realize $b_1\Gl_1^{2m+2}$ and $b_2(\Gl_1+1)^{2m+2}$ for any positive constants
$b_1$ and $b_2$.

Now suppose we are given any polynomial $q(\Gl_1)$ of degree $2m+2$ or less. Using \eq{2.15} we can
express it in the form
\beq q(\Gl_1)=c_1\Gl_1^{2m+2}+c_2(\Gl_1+1)^{2m+2}+r(\Gl_1), \eeq{2.16}
where $r(\Gl_1)$ is of degree $2m$ or less, and the constants $c_1$ and $c_2$ could have either sign, or be zero. Now there exists a sufficiently large constant $c>0$ such that
$c+c_1\Gl_1^{2m+2}$, $c+c_2(\Gl_1+1)^{2m+2}$ and $c+r(\Gl_1)$ are bounded below by a positive constant on  the interval $I$. Hence using the fact that sums or differences of realizable 
functions are realizable (provided the difference is bounded below by a positive constant on $I$) we deduce that each of these three functions is realizable on $I$, and hence too is their
sum $s(\Gl_1)$ in terms of which
\beq q(\Gl_1)=s(\Gl_1)-3c. \eeq{2.17}
Again this is the difference of two realizable functions, and will be realizable if $q(\Gl_1)$ is bounded below by a positive constant on the domain $I$. This proves the assertion is true
for all polynomials of degree $n\leq 2m+2$, and by induction for polynomials of any degree.

Now by the Weierstrass approximation theorem (see, for example, \citeAPY{Keener:2000:PAM}) given any continuous function $f(\Gl_1)$ on $I$, we can approximate
$f(\Gl_1)$ arbitrarily closely by polynomials. Specifically, given any $\Ge>0$ there is a polynomial $p(\Gl_1)$ so that
\beq \max_{\Gl_1\in I}|f(\Gl_1)-p(\Gl_1)|<\Ge. \eeq{2.17a}
Since we can realize $p(\Gl_1)$ arbitrarily closely it follows that we can realize $f(\Gl_1)$ arbitrarily closely.

\subsection{Realizability of any trajectory}
Let $R(\Gl_1^-,\Gl_1^+,\Gl_2^-,\Gl_2^+)$ denote the rectangle 
\beq 0<\Gl_1^-<\Gl_1<\Gl_1^+,\quad 0<\Gl_2^-<\Gl_2<\Gl_2^+, \eeq{2.18}
and let $\CR$ denote the set of all such rectangles. Here we show that any continuous trajectory $(\Gl_1,\Gl_2)=(f_1(t_0),f_2(t_0))$ taking values in a rectangle in $\CR$ is realizable. 
We assume the parameterization has been chosen so $t_0$ has dimensions of length, and $t_0$ increases from
$t_0^->0$ to $t_0^+>t_0^-$ along the trajectory, and 
that $f_1(t_0)$ and $f_2(t_0)$ are defined, bounded, continuous and positive on the closed interval $I=[t_0^-,t_0^+]$. This
implies $f_1(I)\in\cal D$ and $f_2(I)\in\cal D$.   
 The geometry which accomplishes the realizability is that in figure \fig{11}. The blue material, of which there are $n$ cells in the horizontal direction, is a rotation by $90^\circ$ of the material
which realizes $f_1(t_0)$ rescaled so $h=n^{-1}f_1(nt)$.  The red material, of which there are $n$ cells in the vertical direction, is the material
which realizes $f_2(t_0)$ rescaled so $g=n^{-1}f_2(nt)$. We choose $t=t_0/n$  
and we scale the bipod support leg length $d$ as $d=d_0/\sqrt{n}$. Then for $t_0\in I$ we have
\beq nh<\Gl_1<nh+2d=nh+2d_0/\sqrt{n}, \eeq{2.18a}
which implies $f_1(t_0)=nh$ converges to $\Gl_1$ as $n\to\infty$. Similarly  $f_2(t_0)=ng$ converges to $\Gl_2$ as $n\to\infty$. Therefore the trajectory
 $(\Gl_1,\Gl_2)=(f_1(t_0),f_2(t_0))$ is realizable. An interesting corollary is that the trajectory can self-intersect. At points of intersection knowledge of $\Gl_1$ and $\Gl_2$ is 
not sufficient to determine $t_0$ and hence to determine what subsequent deformations are possible: one needs to keep track of the hidden variable $t_0$. 

\begin{figure}[htbp]
\vspace{2.5in}
\hspace{1in}
{\resizebox{2.0in}{1.0in}
{\includegraphics[0in,0in][6in,3in]{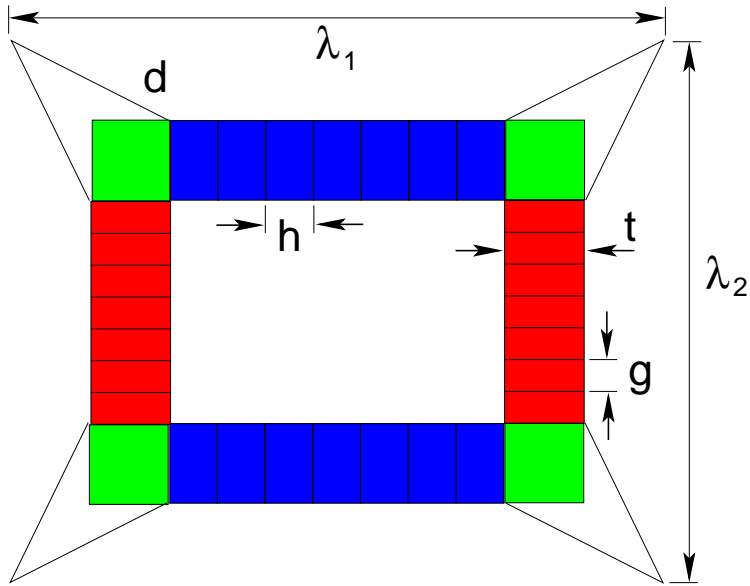}}}
\vspace{0.1in}
\caption{Given two planar rectangular materials (blue and red) having arbitrary response functions one can construct an rectangular 
metamaterial having any desired response trajectory $(\Gl_1(t), \Gl_2(t))$ within a desired domain. Here the green cells are dilator cells. }
\labfig{11}
\end{figure}

\section{Realizability of oblique materials with an arbitrary response}
\setcounter{equation}{0} 

Let us introduce polar coordinates $(r,\Gt)$ such that $(\Gl_1,\Gl_2)=(r\cos\Gt,r\sin\Gt)$. Let $A(r^-,r^+,\Gv)$ denote the sector
\beq 0<r^-<r<r^+,\quad 0<\Gv<\Gt<\pi/2-\Gv, \eeq{2.19}
and let $\CA$ denote the set of all such sectors. Since, with rectangular materials, we are free to realize any trajectory $(\Gl_1(t),\Gl_2(t))$ taking values in a rectangle in $\CR$,
we can in particular realize any trajectory
where $r$ increases strictly monotonically with $t$ such as the blue trajectory in figure \fig{12}. In other words we can realize any continuous desired trajectory $\Gt=\Gf(r)$
taking values in a sector in $\CA$, with $\Gl_1=e_1(r)$ and $\Gl_2=e_2(r)$ for some continuous functions $e_1(r)$ and $e_2(r)$. 

\begin{figure}[htbp]
\vspace{3.0in}
\hspace{1in}
{\resizebox{2.0in}{1.0in}
{\includegraphics[0in,0in][6in,3in]{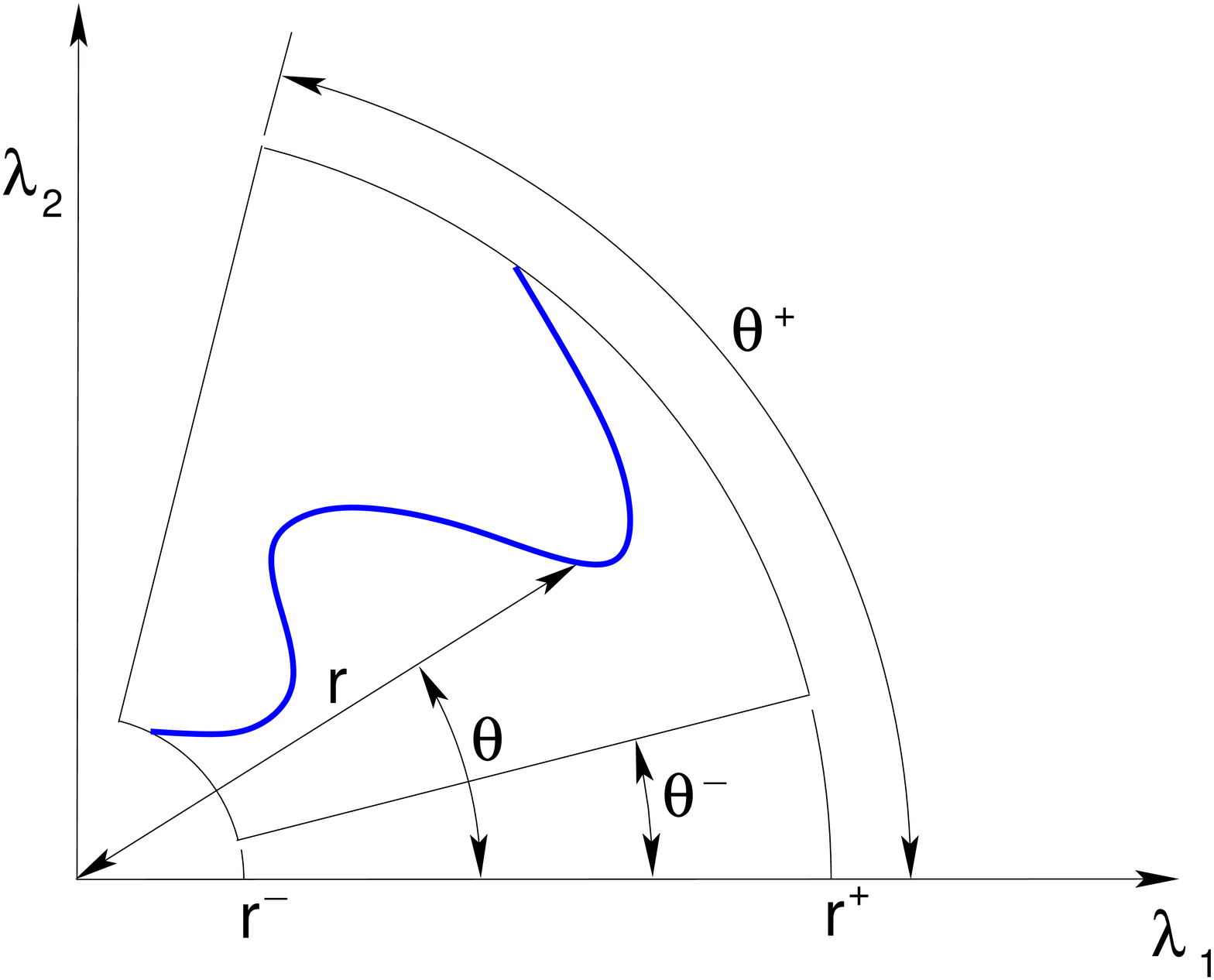}}}
\vspace{0.1in}
\caption{A desired function $\Gt(r)$ taking values in a domain $0<\Gt^-\leq\Gt\leq\Gt^+<\pi/2$, $0<t^-\leq t \leq t^+$ corresponds to a trajectory $(\Gl_1(t), \Gl_2(t))$
(for example the blue one shown here) where $(r,\Gt)$ are the polar coordinates of $(\Gl_1,\Gl_2)$.}
\labfig{12}
\end{figure}

Next consider the geometry of figure \fig{13} which we call an angle adjuster. The yellow cells are rectangular material. Each of the four arms in the cross in is $2$ cells wide and $n$ cells long
(in the figure $n=7$) and the rectangle at the center of the cross is two cells by two cells.  Here by rescaling $h=(2n)^{-1}e_1(2nr)$, $g=(2n)^{-1}e_2(2nr)$ and $\Gt=\Gf(2nr)$ 
in which $r$ is the length of the cell diagonal and $\tan\Gt=g/h$, as shown in the insert. Thus the two horizontal arms have lengths $e_1(2nr)/2$ and the vertical arms have lengths 
$e_2(2nr)/2$. We choose to scale the bipod support leg length $d$ as $d=d_0/\sqrt{n}$. Thus in the limit as $n\to\infty$ the width of the arms shrinks to zero, and the size of the bipod supports
shrink to zero, and $2nr\to t$ while $\Ga\to 2\Gt$. So, by the continuity of $\Gf$, in the limit we have that $\Ga=2\Gf(t)$. Note that by rescaling of the microstructure we can also
realize $\Ga=2\Gf(ct)$ for any constant $c>0$. 

\begin{figure}[htbp]
\vspace{3in}
\hspace{0.2in}
{\resizebox{2.0in}{1.0in}
{\includegraphics[0in,0in][8in,4in]{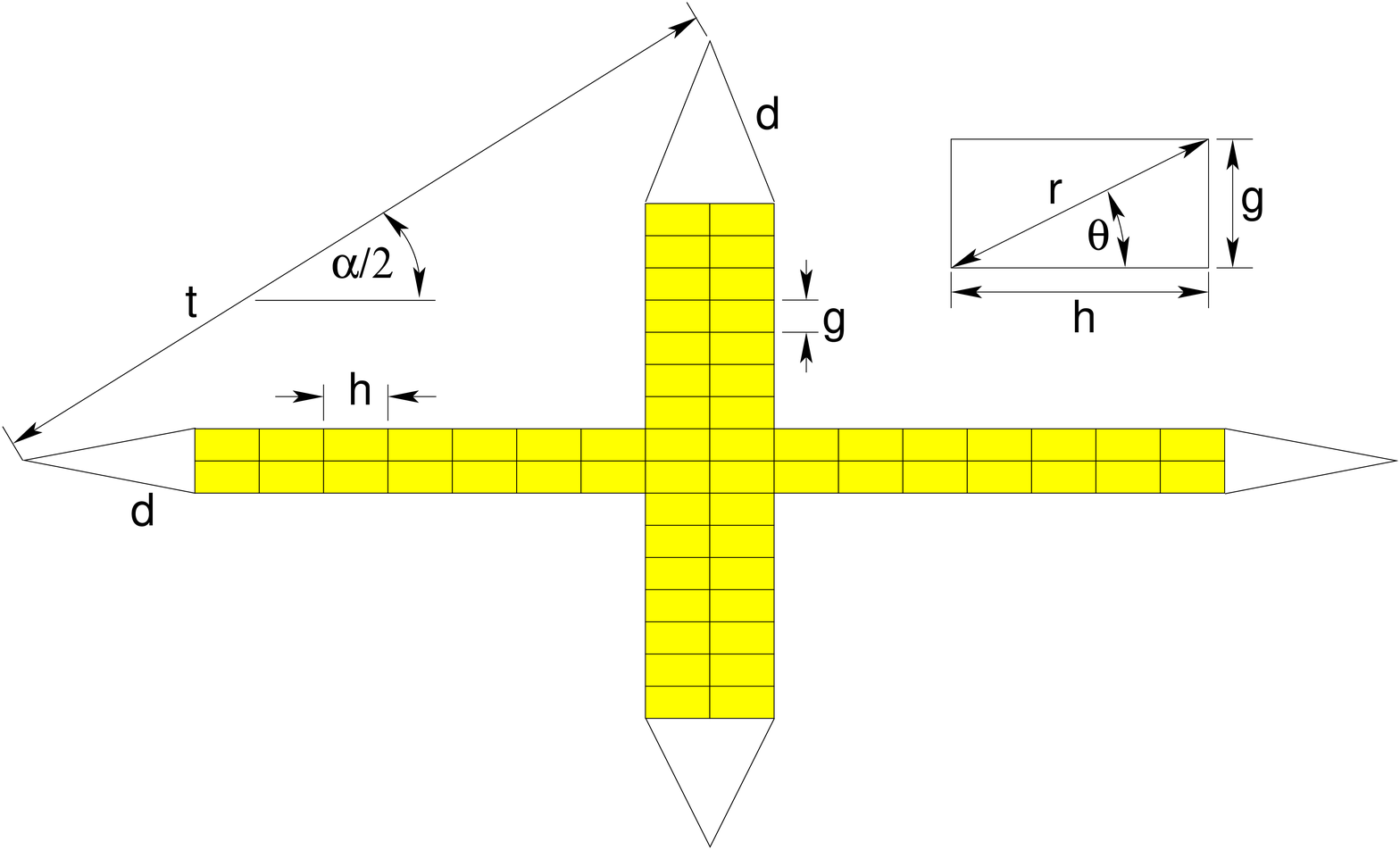}}}
\vspace{0.1in}
\caption{By taking a material (shown yellow here) with an rectangular response one can obtain an angle adjuster with a cell which is rhombus. The insert shows the geometry of one cell.}
\labfig{13}
\end{figure}

With this angle adjuster we can now obtain oblique materials (i.e. periodic materials with an underlying oblique lattice) 
with an arbitrary response through the microstructure of figure \fig{14}. Now $\Bu$ and $\Bv$ are no 
longer orthogonal, and we let $\Gj$, with $0<\Gj<\pi$ denote the angle between these vectors, which varies as the cell deforms.
In the corners is one cell (yellow) which is an angle adjuster. 
On the top and bottom there are $n$ cells of blue rectangular material and on the sides there are $n$ cells of red rectangular material. 
We define $\cal B$ as the set of all finite connected subintervals contained strictly in $[0,\pi]$, which may be open or closed at either end:
\beq \CB =\{ (a,b),(a,b],[a,b),[a,b]~|~0<a<b<\pi \} . \eeq{2.19aa}
Suppose we are given any two functions $f_1(t_0)$ and $f_2(t_0)$ which are continuous on an open interval $D\in \cal D$, with $f_1(D)\in\cal D$ and $f_2(D)\in\cal D$, 
and a function $\Gf(t_0)$ which is continuous on $D$ with $\Gf(D)\in\cal B$. We can, by rescaling
realize a blue rectangular cell with $h=n^{-1}f_1(nt)$, a red rectangular cell with $h=n^{-1}f_2(nt)$ and a yellow angle adjuster cell with $\Ga=\Gf(nt)$ for $nt$ in any closed subinterval 
$I=[t_0^-,t_0^+]\subset D$. We choose $t=t_0/n$
and scale the bipod support leg length $d$ as $d=d_0/\sqrt{n}$. Then in the limit $n\to\infty$ we have $\Gl_1\equiv|\Bu|\to nh$, $\Gl_2\equiv|\Bv|\to ng$ and $\Ga\to \Gj$. In this limit it then follows that 
$\Gl_1=f_1(t_0)$, $\Gl_2=f_2(t_0)$ and $\Gj=\Gf(t_0)$ for all $t_0\in I$. We have thus realized
any desired response. 

Finally, to make sure the trajectory does not extend outside $[t_0^-,t_0^+]$ we choose the dilators at the corners in figure \fig{14} so that their performance range
is exactly $nt\in(t_0^-,t_0^+)$ or, alternatively, given the dilator performance range we parameterize the trajectory to match it. The endpoints
of the trajectory $t_0=t_0^-$ and $t_0=t_0^+$ might then not be realizable.

\begin{figure}[htbp]
\vspace{3.0in}
\hspace{0.2in}
{\resizebox{2.0in}{1.0in}
{\includegraphics[0in,0in][7in,3.5in]{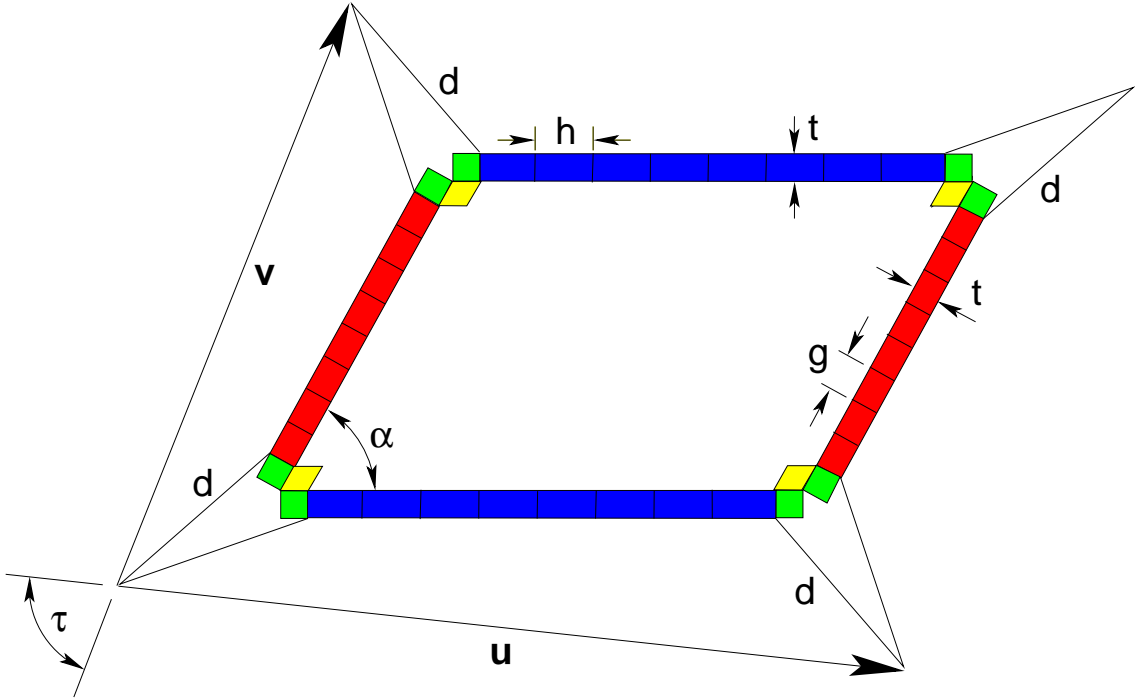}}}
\vspace{0.1in}
\caption{By combining two rectangular materials (blue and red) and a rhombus shaped angle adjuster (yellow) one can obtain a 
oblique material having an arbitrary response. The green cells near the corners are dilator cells. }
\labfig{14}
\end{figure}

\section{Realizability of three-dimensional orthorhombic materials with an arbitrary response}
\setcounter{equation}{0} 

A key element in the construction of three dimensional materials with any desired response is the construction of panels from two dimensional rectangular or parallelogram cells lying in 
the $x_1-x_2$
plane. An example of the construction of a panel from the expander cell is given in figure \fig{15}(a). The first step is to extend the microstructure a short distance $r$ (short compared to the cell
size) in the positive $x_3$ direction, so rigid lines become rigid faces, rigid triangles become rigid triangular prisms, rectangles become rectangular prisms, and so forth. 
Pivot junctions become edge junctions. The next step is to modify the structure so it tapers to a point at the four supports. The final step is the bend the tips of these 
supports a short distance $s$ in the negative $x_3$ direction, so that the microstructure is strictly contained in a box having the four supports as corners of one face. The final construction
may have rigid flat rectangles as in \fig{15}(b).  If one a desires a construction that only involves rigid bars then one should first make the rectangle into a thin triangular prism as
shown in figure \fig{15}(c) and then make the transformation to a truss of bars as in figure \fig{0a}(b).

\begin{figure}[htbp]
\vspace{2in}
\hspace{1in}
{\resizebox{2.0in}{1.0in}
{\includegraphics[0in,0in][6in,3in]{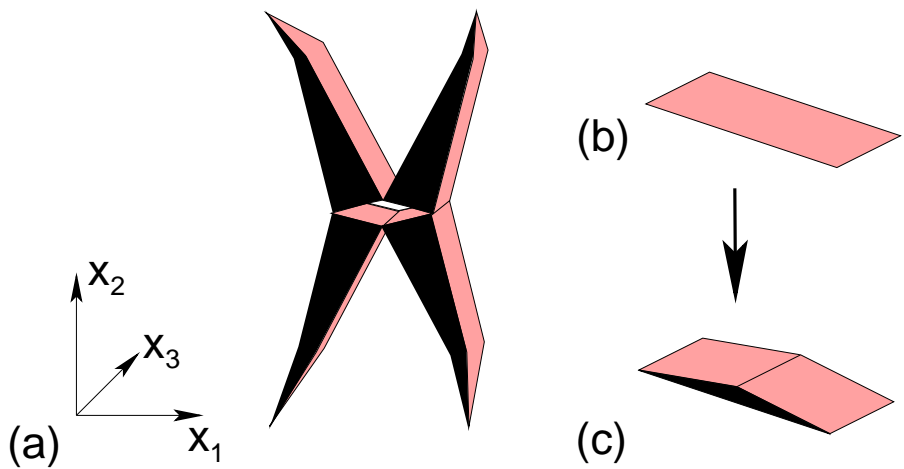}}}
\vspace{0.1in}
\caption{As shown in (a) two dimensional cells can be extended into the $x_3$-direction to obtain panels, which can be used to construct three dimensional cells. The panels
are tapered at their four supports, and the tapered ends are bent slightly in the $x_3$-direction, so that all the microstructure is to one side of the 
plane through the four support points. To avoid confusion those
faces of the rigid objects parallel to the $x_3$-axis are colored in pink. Any rigid flat rectangles in the final construction as in (b) can be transformed
into thin triangular prisms as in (c) and then replaced by a truss of bars if desired.}
\labfig{15}
\end{figure}

Following the construction of \citeAPY{Buckmann:2012:RTD} and as shown in figure \fig{16} 
one can assemble 5 (or 6) panels based on the cell geometry of figure \fig{4} to form a cubic cell which acts as a three-dimensional dilator. 
To obtain three-dimensional dilator cells which allow arbitrary expansion one could use the cell structure of figure \fig{5} as the basis for the panels. 
A periodic dilational  material of these cubic cells would have an
arbitrarily large flexibility window as defined by \citeAPY{Sartbaeva:2006:FWZ}.

\begin{figure}[htbp]
\vspace{2in}
\hspace{0.1in}
{\resizebox{2.5in}{1.25in}
{\includegraphics[0in,0in][6in,3in]{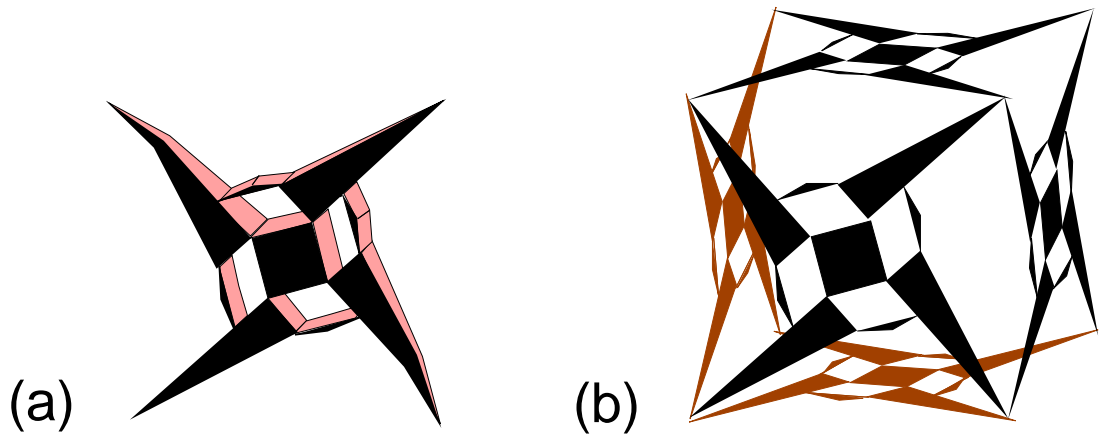}}}
\vspace{0.1in}
\caption{Shown in (a) is the panel which corresponds to the two-dimensional dilator cell of figure \fig{4}. To avoid confusion those
faces of the rigid objects parallel to the $x_3$-axis are colored in pink. Following \protect\citeAPY{Buckmann:2012:RTD}
copies of these can be assembled as the
faces of a cube to form a three-dimensional dilator cell, as illustrated in (b). Only five, rather than six, panels are shown. In fact these are all that are needed. For clarity 
two of these panels are colored in brown and the three dimensional structure of each panel is omitted. They are oriented so the microstructure lies strictly inside
the cube formed by the eight support points.}
\labfig{16}
\end{figure}

Now we can use these dilators to construct three-dimensional orthorhombic materials with an arbitrary response. By orthorhombic we mean that the vectors $\Bu$, $\Bv$ and $\Bw$ are mutually
orthogonal and remain so as the material deforms. Let the cell of the material, that we will construct, have sides of
lengths $\Gl_1\equiv|\Bu|$, $\Gl_2\equiv|\Bv|$ and $\Gl_3\equiv|\Bw|$ in the $x_1$, $x_2$, and $x_3$ directions. Let $P(\Gl_1^-,\Gl_1^+,\Gl_2^-,\Gl_2^+, \Gl_3^-,\Gl_3^+)$ denote the rectangular prism 
\beq 0<\Gl_1^-<\Gl_1<\Gl_1^+,\quad 0<\Gl_2^-<\Gl_2<\Gl_2^+, \quad 0<\Gl_3^-<\Gl_3<\Gl_3^+,  \eeq{2.19a}
and let $\CP$ denote the set of all such rectangular prisms.  Here we show that any trajectory $(\Gl_1,\Gl_2,\Gl_3)=(f_1(t_0),f_2(t_0),f_3(t_0))$ taking values in a rectangular prism in $\CP$ is realizable. 
We assume the parameterization has been chosen so $t_0$ has dimensions of length, and $t_0$ increases from
$t_0^->0$ to $t_0^+>t_0^-$ along the trajectory, and that $f_1(t_0)$, $f_2(t_0)$, and $f_3(t_0)$ are defined, bounded, continuous and positive on the closed interval $I=[t_0^-,t_0^+]$. This
implies $f_1(I)\in\cal D$, $f_2(I)\in\cal D$ and $f_3(I)\in\cal D$. The corner of    
the geometry which accomplishes the realizability is shown in figure \fig{17}. This geometry is a generalization
of the two dimensional geometry of figure \fig{11}. In the $x_1$, $x_2$, and $x_3$ directions there are blue, red, and cyan tubes, respectively, which are each $n$ units long, 
where each unit consists of four panels joined to make a square tube. Each tube retains its square cross section due to its contact with the green dilator cells.  
Choose $t=t_0/n$, where $t_0\in I$, and scale the tripod support leg length $d$ as $d=d_0/\sqrt{n}$.
From the results of section 2.4 and rescaling we can realize $h=n^{-1}f_1(nt)$, $g=n^{-1}f_2(nt)$ and $s=n^{-1}f_3(nt)$. Then for $t_0\in I$ we have
\beq nh<\Gl_1<nh+2d=nh+2d_0/\sqrt{n}, \eeq{2.19b}
which implies $f_1(t_0)=nh\to\Gl_1$ as $n\to\infty$. Similarly  it follows that $f_2(t_0)=ng\to\Gl_2$ and $f_3(t_0)=ns\to\Gl_3$ as $n\to\infty$.
Therefore the trajectory $(\Gl_1,\Gl_2,\Gl_3)=(f_1(t_0),f_2(t_0),f_3(t_0))$ is realizable.

\begin{figure}[htbp]
\vspace{3.3in}
\hspace{0.5in}
{\resizebox{2.5in}{1.5in}
{\includegraphics[0in,0in][6in,3in]{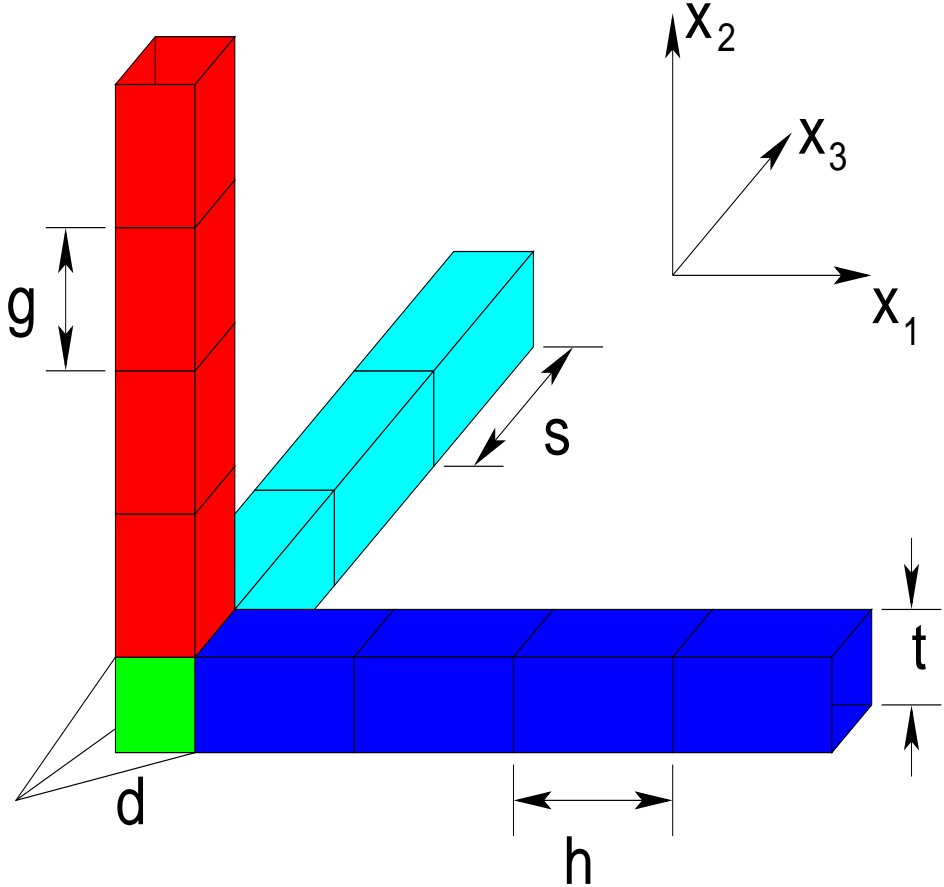}}}
\vspace{0.1in}
\caption{Corner of the cell which realizes any desired orthorhombic response. The green cell is a three-dimensional dilator cell and the blue, red, and cyan tubes, with square
cross section, have panels based on cells with the needed response.}
\labfig{17}
\end{figure}

\section{Realizability of three-dimensional triclinic materials with an arbitrary response}
\setcounter{equation}{0} 

The construction of three-dimensional triclinic materials with an arbitrary response, is similar in some respects to the two-dimensional construction of figure \fig{14}. We let
$\Gf_1$ denote the angle between $\Bu$ and $\Bv$, $\Gf_2$ the angle between $\Bv$ and $\Bw$, and $\Gf_3$ the angle between $\Bw$ and $\Bu$. 
Assuming the vectors $\Bu$, $\Bv$ and $\Bw$ are not coplanar (since otherwise $\BA=\BF^T\BF$ is singular) these angles satisfy
\beq 0<\Gf_1<\pi,\quad 0<\Gf_2<\pi,\quad 0<\Gf_3<\pi,\quad \Gf_1+\Gf_2+\Gf_3<2\pi. \eeq{2.19c}
The first step is the construction of a three-dimensional angle adjuster as shown in figure \fig{18}(a), just before the points $D_1$ and $D_2$ are joined. The construction of one corner of the cell
then proceeds as shown in figure \fig{18}(b). See the caption for more details. In the $x_1$, $x_2$, and $x_3$ directions there are blue, red, and cyan tubes, respectively, which are each $n$ units long, 
where each unit consists of four panels joined to make a square tube. Three additional rhombi, with side length $mt$ 
(composed of  $m\times m$ orange, $m\times m$ magenta, and $m\times m$ gold panels) can be added to 
the corner of the cell of figure \fig{18}(b)
to form a rhombohedron, which can be copied $7$ times and translated to the other $7$ corners of the cell and joined to the tubes with dilator cells, in similar
way to the construction of \fig{18}(b). Now if one has two line segments of length $nh+2(m-1)t$ (the length of the interval between $C$ and the equivalent point in the vertical direction at the
top corner of the rhombohedron above it) joined to two line segments of length $ng+2(m-1)t$ where the angles between line segments are either $\Ga$ or $\pi-\Ga$, then those line
segments must form a parallelogram (assuming, as is the case here, that ``opposite angles are the same'' to avoid the case of a warped isosceles trapezoid). 
The corresponding line segments from all sides form a parallelepiped, with angles of $\Ga$, $\pi-\Ga$, $\Gb$, $\pi-\Gb$, $\Gg$, and $\pi-\Gg$, with side
lengths of $nh+2(m-1)t$, $ng+2(m-1)t$, and $ns+2(m-1)t$. A distance $t$ from the corners are the attachments to the tripods, having leg length $d$. 
So the support is at most a distance $d$ from these attachments. Thus we have the inequalities
\beqa nh+2(m-2)t-d< & \Gl_1 & <nh+2(m-2)t+d,\nonum
      ng+2(m-2)t-d< & \Gl_2 & <ng+2(m-2)t+d, \nonum 
      ns+2(m-2)t-d< & \Gl_3 &<ns+2(m-2)t+d,
\eeqa{2.20}
where $\Gl_1\equiv |\Bu|$, $\Gl_2\equiv |\Bv|$, and $\Gl_3\equiv |\Bw|$ are the lengths of the sides of that parallelepiped which has the eight support points (each at the end of a tripod) as vertices: see
figure \fig{19}.

Next suppose we are given any three functions $f_1(t_0)$, $f_2(t_0)$ and $f_3(t_0)$ 
which are continuous on an open interval $D\in \cal D$, with $f_i(D)\in\cal D$ for $i=1,2,3$, and three functions
 $\Gf_1(t_0)$, $\Gf_2(t_0)$ and $\Gf_3(t_0)$ which are continuous on $D\in \cal D$, with $\Gf_i(D)\in\cal B$ for $i=1,2,3$,
and additionally with $\Gvf(D)\in\cal B$ where
\beq \Gvf(t_0)=(\Gf_1(t_0)+\Gf_2(t_0)+\Gf_3(t_0))/2. \eeq{2.21}
We can realize, by rescaling,
\begin{itemize}
\item a blue rectangular panel with $h=n^{-1}f_1(nt)$, 
\item a red rectangular panel with $g=n^{-1}f_2(nt)$ 
\item a cyan rectangular panel with $s=n^{-1}f_2(nt)$
\item a orange angle adjuster panel with $\Ga=\Gf_1(nt)$
\item a magenta angle adjuster panel with $\Gb=\Gf_2(nt)$
\item a gold angle adjuster panel with $\Gg=\Gf_3(nt)$
\end{itemize}
for $nt$ in any closed subinterval $I=[t_0^-,t_0^+]\subset D$.
We choose $t=t_0/n$ and scale the tripod support leg length $d$ as $d=d_0/\sqrt{n}$. Since $\Gf_i(D)\in\cal B$ for $i=1,2,3$ it follows that there exists an angle $\Gn$
such that 
\beq \Gn<\Gf_i(t_0)<\pi-\Gn, \quad i=1,2,3. \eeq{2.22}
Thus, for $t_0\in I$ none of the angles $\Ga$, $\pi-\Ga$, $\Gb$, $\pi-\Gb$, $\Gg$, and $\pi-\Gg$ are less than $\Gn$.  We choose $m$ independent of $n$ and such
that 
\beq m>1+1/\tan(\Gn/2), \eeq{2.23}
to ensure that none of the green dilation cells touch one another (and possibly jam the structure) as the cell deforms: see figure \fig{20}.
Then in the limit $n\to\infty$ \eq{2.20} implies
 $\Gl_1\to nh$, $\Gl_2\to ng$ and $\Gl_3\to ns$. Likewise, because the size of the supports shrinks to zero, 
$\Gj_1\to \Ga$, $\Gj_2\to \Gb$ and $\Gj_3\to\Gg$. This, implies that in the limit $n\to\infty$,
\beq \Gl_i=f_i(t_0), \quad \Gj_i=\Gf_i(t_0), \quad i=1,2,3.
\eeq{2.24} 
We have thus realized any desired triclinic response. Finally, to make sure the trajectory does not extend outside $[t_0^-,t_0^+]$ 
we choose the dilators in figure \fig{18}(b) so that their (rescaled) performance range
is exactly $nt\in (t_0^-,t_0^+)$ or, alternatively, given the dilator performance range we parameterize the trajectory to match it. The endpoints
of the trajectory $t_0=t_0^-$ and $t_0=t_0^+$ might then not be realizable. If an actuator is placed within one of these dilator cells in the material
then it should be possible to adjust $t_0$ to any desired position along the trajectory.

\begin{figure}[htbp]
\vspace{4in}
\hspace{0in}
{\resizebox{2.0in}{1.0in}
{\includegraphics[0in,0in][6in,3in]{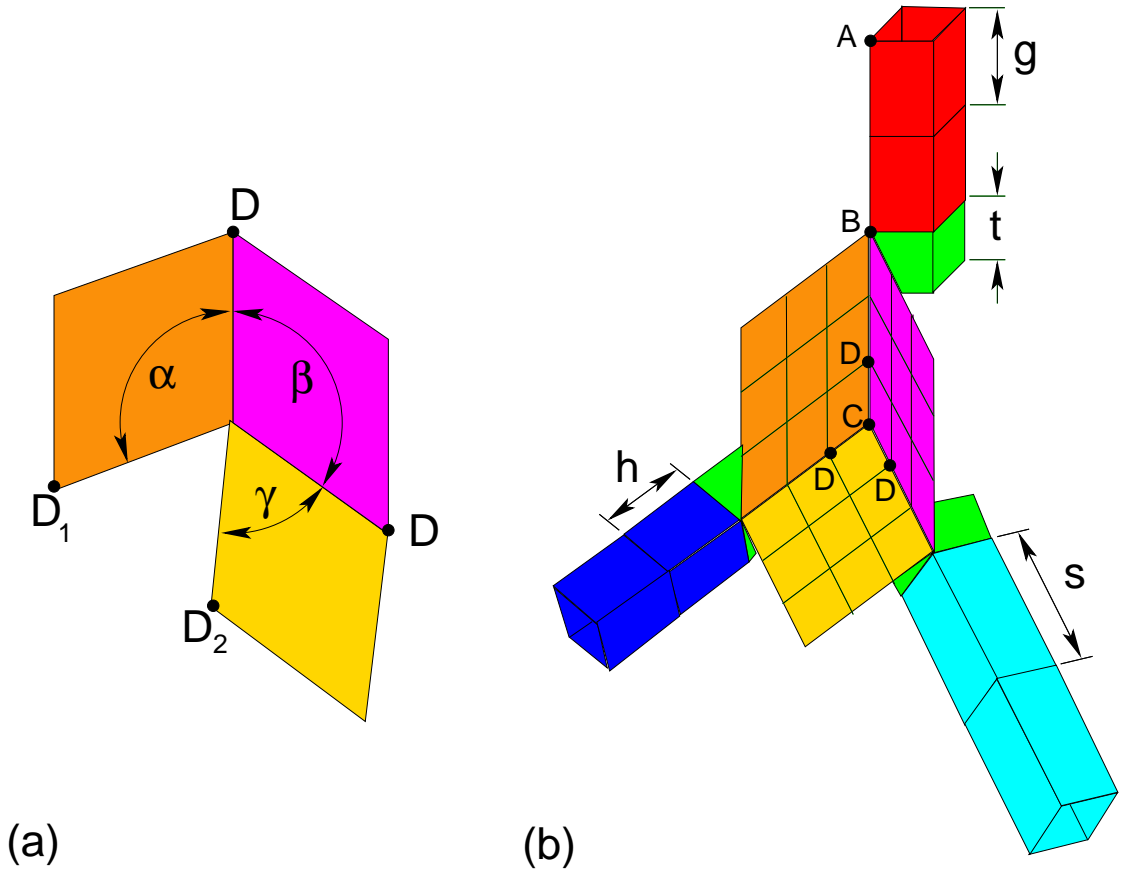}}}
\vspace{0.1in}
\caption{(a) The cells of three planar angle adjusters are made into panels (colored orange, magenta, and gold) and attached at their supports, as shown here in planar configuration before the points
$D_1$ and $D_2$ are joined. Note that $\Ga+\Gb+\Gg$ must be less than $2\pi$ which accounts for the condition $\Gvf(D)\in\cal B$.
(b) Corner of the cell which realizes any desired triclinic response. The points $A$, $B$ and $C$ are colinear in the
correct visualization, and the point $C$ is behind the plane going through the three points labelled $D$: thus 
the perspective is like that obtained by looking from inside a room towards a corner on the floor.   The green cells are three-dimensional dilator cells and the attached 
blue, red, and cyan tubes, with square cross section, have panels based on rectangular cells with the needed response. 
There are $m\times m$ panels of each of the three angle adjusters (orange, magenta, and gold), joined to form three rhombi. 
(In each rhombus the supports at the $(m-1)^2$ interior points need not be tapered). The figure has $m=3$.
One green dilator cell has an edge in common with those orange and magenta panels that touch the point
$B$. The green dilator cell and the attached red tube is free to rotate in the plane perpendicular to this edge. The two other green dilator cells are attached in a similar way. A tripod support (not shown,
as it is behind the corner) has its three legs, each of length $d$, attached to the three points marked $D$. }
\labfig{18}
\end{figure}

\begin{figure}[htbp]
\vspace{3.5in}
\hspace{2in}
{\resizebox{2.0in}{1.0in}
{\includegraphics[0in,0in][8in,4in]{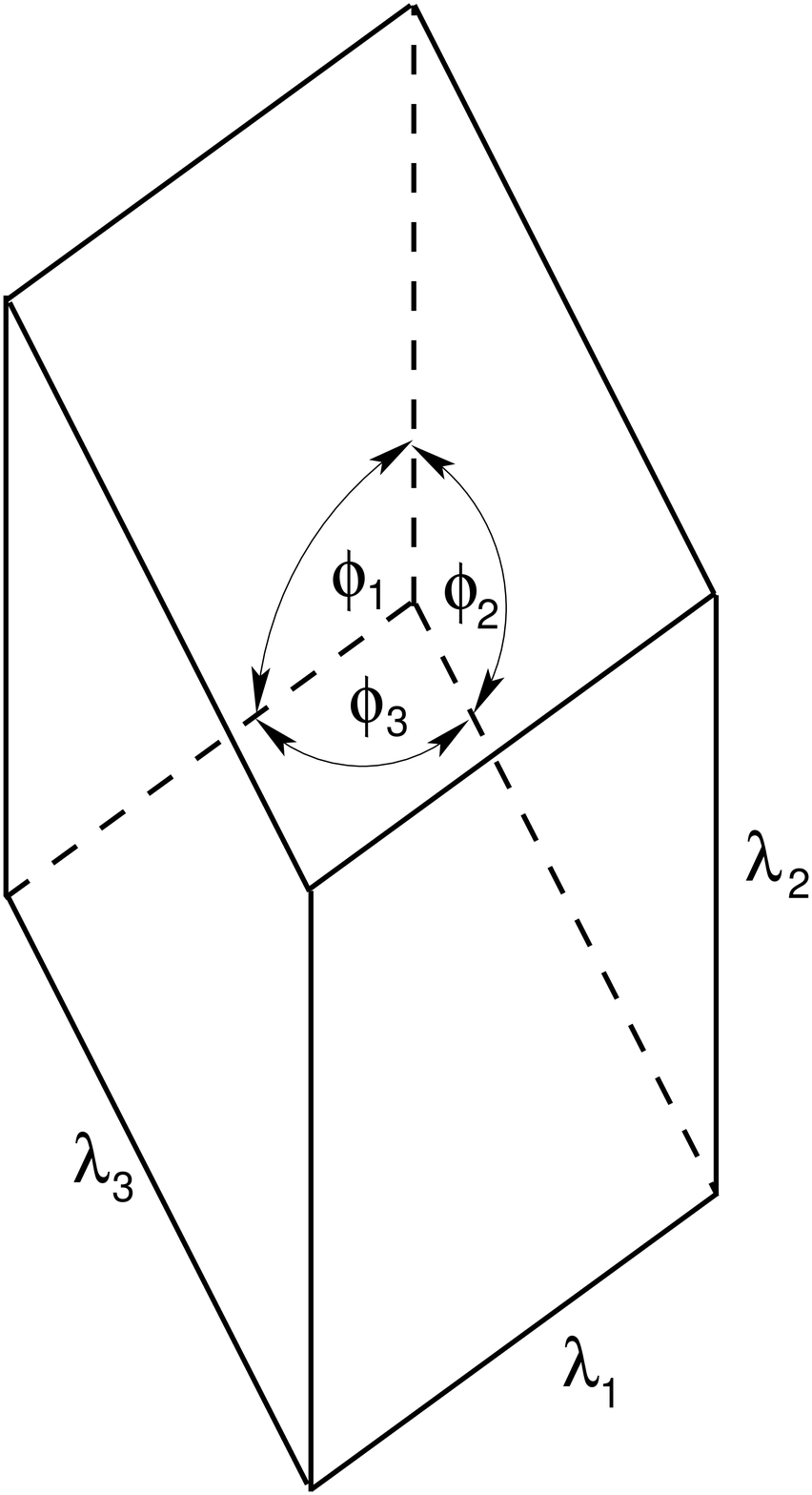}}}
\vspace{0.1in}
\caption{The vertices of this parallelepiped are the support points, each at the end of a tripod. The microstructure inside the cell, including the tripods, is not shown, but can be
visualized using figure \fig{18}: that corner corresponds to the corner here where the three dashed lines meet, although the angles are slightly different due to the presence of the tripods.
The angles $\Gf_1$, $\Gf_2$ and $\Gf_3$ are the actual angles between lines (not the angles between their projections) and thus sum to less than
$360^{\circ}$. When they are all $90^\circ$ the parallelepiped becomes a rectangular prism.}
\labfig{19}
\end{figure}

\begin{figure}[htbp]
\vspace{1in}
\hspace{1in}
{\resizebox{2.0in}{1.0in}
{\includegraphics[0in,0in][6in,3in]{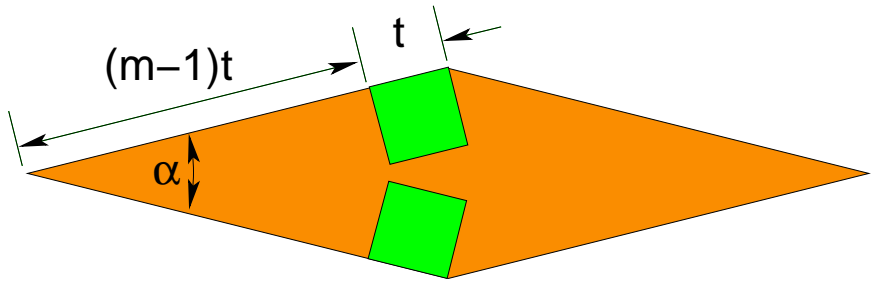}}}
\vspace{0.1in}
\caption{The worst configuation of the green dilator cells relative to an orange angle adjuster. They avoid collision if $\tan(\Ga/2)>1/(m-1)$. Hence follows the condition \eq{2.23}. }
\labfig{20}
\end{figure}
\section*{Acknowledgements}
GWM is grateful to the reviewer for helpful comments and is thankful for support from the National Science Foundation through grant DMS-1211359.

\bibliography{/home/milton/tcbook,/home/milton/newref}
\bibliographystyle{/home/milton/latex/unimode/mod-xchicago}
\end{document}